\documentclass[authoryear,preprint,review,12pt]{elsarticle}
\usepackage{amsmath, amsthm, amssymb}
\usepackage{array}
\usepackage{fancybox,amssymb}
\usepackage[dvips]{color}
\usepackage{rotating}
\usepackage{bm}
\usepackage{tabularx}
\usepackage{float}
\usepackage{natbib}
\usepackage{subfigure}
\usepackage{graphicx}
\usepackage{epstopdf}
\usepackage{booktabs}
\usepackage{pdflscape}
\usepackage{appendix}
\usepackage[linesnumbered,ruled]{algorithm2e}
\allowdisplaybreaks
\newcommand{\Var}{{\rm Var}}
\newcommand{\E}{{\rm E}}

\newtheorem{theorem}{Theorem}

\def\5n{\negthinspace \negthinspace \negthinspace \negthinspace \negthinspace }
\def\4n{\negthinspace \negthinspace \negthinspace \negthinspace }
\def\3n{\negthinspace \negthinspace \negthinspace }
\def\2n{\negthinspace \negthinspace }

\def\0{\mathbf{0}}
\def\1{\mathbf{1}}
\def\E{\mathbb{E}}

\def\H{\mathcal{H}}
\def\R{\mathbb{R}}

\begin{document}

\begin{frontmatter}

\title{Discrete-Time Mean-Variance Strategy Based on Reinforcement Learning}

\author[a]{Xiangyu Cui\corref{cor1}\fnref{fn1}}
\ead{cui.xiangyu@mail.shufe.edu.cn}
\address[a]{School of Statistics and Management, Shanghai University of Finance and Economics.}

\author[b]{Xun Li}
\ead{li.xun@polyu.edu.hk}
\address[b]{Department of Applied Mathematics, The Hong Kong Polytechnic University}

\author[c]{Yun Shi}
\ead{yshi@fem.ecnu.edu.cn}
\address[c]{School of Statistics and Academy of Statistics and Interdisciplinary Sciences, East China Normal University}

\author[a]{Si Zhao}
%\ead{zhaosi19980613@163.com}

\cortext[cor1]{Corresponding author.}
\fntext[fn1]{Postal address: 777 Guoding Rd., Shanghai,200433, P. R. China.}

\begin{abstract}
This paper studies a discrete-time mean-variance model based on reinforcement learning. Compared with its continuous-time counterpart in \cite{zhou2020mv}, the discrete-time model makes more general assumptions about the asset's return distribution. Using entropy to measure the cost of exploration, we derive the optimal investment strategy, whose density function is also Gaussian type. Additionally, we design the corresponding reinforcement learning algorithm. Both simulation experiments and empirical analysis indicate that our discrete-time model exhibits better applicability when analyzing real-world data than the continuous-time model. 
\end{abstract}

\end{frontmatter}

\section{Introduction}
The mean-variance portfolio selection model, first proposed by \cite{Markowitz1952}, has long been recognized as the cornerstone of modern portfolio theory. \cite{Markowitz1952} built a single-period framework to investigate the portfolio selection problem, where an investor seeks a portfolio to maximize the expected total return for any given level of risk measured by variance. The mean–variance framework is of particular interest because it not only captures the trade-off between portfolio return and risk but also suffers from the time-inconsistency problem. \cite{li2000} first made a breakthrough and derived the analytical solution to the discrete-time multi-period mean-variance problem. They applied an embedding approach, transforming the mean–variance problem into an LQ problem where classical approaches can be used to find the solution. \cite{zhou2000} adopted the same approach to solve the continuous-time mean-variance problem. \cite{li2002} and \cite{cui2014} tackled` the mean-variance portfolio selection problem under no-shorting constraints in the continuous-time and discrete-time settings, respectively. The MV problem has also been investigated in the hedging by \cite{Duffie1991} and the optimal liquidation by \cite{Almgren2001} among many other variants. \cite{cui2022} provided a comprehensive survey on the multi-period mean–variance portfolio selection model.

The classical stochastic control approach for solving portfolio optimization across multiple assets requires representations of the dynamics of individual assets and their co-movements, which are difficult to describe and estimate accurately. Applications of reinforcement learning to finance, such as high-frequency trading and portfolio management, have attracted much attention in recent years. RL describes methods by which agents learn to make optimal decisions through interacting with the environment. \cite{hambly2023} provided a comprehensive review of the recent advances in reinforcement learning in finance and discussed in detail the applications of these RL algorithms in various decision-making problems in finance. 

The portfolio optimization problem can be reformulated as a discrete-time Partially Observable Markov Decision Process (POMDP). \cite{maringer2012} and \cite{maringer2014} presented the regime-switching recurrent reinforcement learning model (RSRRL), a combined technique of statistical modeling and machine learning, to tackle investment problems. \cite{asiain2018} provided a reinforcement learning framework for computing the mean-variance customer portfolio in POMDPs with transaction costs. \cite{liu2020} formulated the quantitative trading process as a POMDP and presented a model based on deep reinforcement learning and imitation learning techniques. \cite{Mannor2013} concluded that seeking the global optimum for MDP problems under the MV criterion is computationally challenging. The computational complexity can be attributed to two reasons: the absence of a principle of optimality leading to simple recursive algorithms and the variance, which is not a linear function of the probability measure of the underlying process. According to \cite{Sato2019}, in the portfolio optimization, neither the future returns of investments nor the state transition probabilities are known. Thus, model-free RL methods can be applied to the problem because one can solve a Bellman optimality equation approximately without understanding the underlying dynamics but relying solely on the sample data.  

Model-free RL methods can be divided into two mainstreams, including value-based and policy-based methods. In the value-based methods, such as Q-Learning, the agent's action is determined not by finding the optimal policy $\boldsymbol{\pi^*}$ directly but instead by obtaining the optimal state-action value function $Q^*(s, a)$. \cite{Du2016} used Q-Learning without a neural network in discretized market states to optimize a portfolio of a riskless asset and a risky asset with transaction costs being taken into account. \cite{Jin2016} and \cite{Weijs2018} tackled the portfolio optimization problem by using Q-Learning with a neural network to approximate the state-action value function. Other value-based methods, including a portfolio trading strategy based on DQN proposed by \cite{Park2020} and a G-Learning-based algorithm established by \cite{Dixon2020}, are also applied to solve the wealth management problem. On the other hand, in the policy-based methods, the policy is directly parameterized with $\boldsymbol{\theta}$ to represent the optimal policy $\boldsymbol{\pi}^* \sim \boldsymbol{\pi}_{\boldsymbol{\theta}}$. \cite{Jiang2017} proposed a framework combining neural networks with DPG and used continuous state inputs to approximate the optimal policy directly. DDPG proposed by \cite{lillicrap2015} is a model-free, off-policy, and actor-critic method combining DPG and DQN. \cite{Liu2018} explored the DDPG algorithm for the portfolio selection of 30 stocks. \cite{liang2018} implemented three RL algorithms, including DDPG, PPO, and PG, under an adversarial training method in portfolio management problems. Applying model-free RL methods to search for the global optimum under the MV criterion in the portfolio optimization problem is still appealing.

In addition to the abovementioned methods, \cite{Wang2020Reinforcement} developed a general entropy-regularized relaxed stochastic control formulation, called an exploratory formulation, to capture the trade-off between exploration and exploitation in RL. Based on the foundation laid by \cite{Wang2020Reinforcement}, \cite{zhou2020mv} established the entropy-regularized continuous-time mean–variance framework with one risky asset and one risk-free asset. They proved that the optimal policy is Gaussian with decaying variance. Also, they proposed an Exploratory Mean-Variance (EMV) algorithm consisting of three procedures: policy evaluation, policy improvement, and a self-correcting scheme for learning the Lagrange multiplier. They showed that this EMV algorithm outperforms two benchmarks, including the analytical solution with estimated model parameters obtained from MLE and DDPG algorithms. \cite{wang2019} generalized the continuous-time framework to large-scale portfolio selection setting with $d$ risky assets and one risk-free asset. He showed that the performance of the EMV algorithm on price data from the stocks in the S\&P 500 with $d \geq 20$ is better than several methods, including DDPG. \cite{jia2022-1}, \cite{jia2022-2} and \cite{jia2023} extended the theory of \cite{Wang2020Reinforcement} from policy evaluation, policy gradient and q-learning. \cite{dai2023} extended the exploratory stochastic control framework to an incomplete market and learned an equilibrium policy under an MV criterion. \cite{wu2023} extended the work of \cite{zhou2020mv} to solve the continuous-time mean-variance portfolio selection problem in a regime-switching market.

Motivated by the work in \cite{zhou2020mv}, we establish an RL framework for studying the solution to the discrete-time MV portfolio selection problem. The first contribution of this paper is to present the global optimal solution to the discrete-time exploratory MV problem. We also find that the optimal feedback control policy is Gaussian with a time-decaying variance, which suggests that the level of exploration decreases as time passes. Another interesting result is the perfect separation between exploitation and exploration in the mean and variance of the optimal Gaussian policy.

Another important contribution of this paper is the design of an effective RL algorithm to find the global optimal solution to the discrete-time exploratory MV problem, premised upon the policy improvement and policy convergence theorem. This theorem provides an updating scheme for the feedback Gaussian policy. It guarantees the fast convergence of both the policy and the value function to the global optimum of the discrete-time EMV problem.

We then test our RL algorithm via simulation and empirical analyses. In the simulation study, we compare our method with the EMV algorithm proposed by \cite{zhou2020mv}. The comparisons are performed under stationary simulated market scenarios, where excess returns of the risky asset are subjected to the skewed $T$ distribution with fixed mean and volatility parameters. Our algorithm achieves better out-of-sample performances than the EMV algorithm in \cite{zhou2020mv} in all simulations. We backtest our algorithm using the S\&P 500 index data in the empirical analysis. We also find that our algorithm brings better out-of-sample returns and converges relatively faster than the EMV algorithm in \cite{zhou2020mv}

An RL framework established to solve the MV problem is always based on basic financial market assumptions. \cite{zhou2020mv} assumed that the price process of the risky asset is a geometric Brownian motion with $S_0 = s_0 > 0$ being the initial price at $t = 0$. In this paper, we want to highlight that without the assumption that the risky asset's price process is a geometric Brownian motion, the EMV algorithm designed by \cite{zhou2020mv} may not work very well. Weaker and more general assumptions about the financial markets on which an RL framework is based may enhance the model's applicability. If the model describing the financial markets deviates far from the conditions in the real world, the output policy won't achieve great investment performance.

The rest of the paper is organized as follows. In Section 2, we present the discrete-time exploratory MV problem and derive the optimal solutions and value functions of this problem. We then prove the policy improvement and convergence result in Section 3. In Sections 4 and 5, we carry out simulation and empirical studies, respectively. Finally, we make brief conclusions in Section 6.

\section{Discrete Time Exploratory Mean-Variance Problem}
\subsection{Classic discrete-time mean-variance problem}
We first recall the classic discrete-time mean-variance problem. We consider a market consisting of two assets: one risk-free asset and one risky asset. The return rate of a risk-free asset is $r_f$ and $r_t$ denotes the excess return of risky asset from period $t$ to period $t+1$, which subjects to a normal distribution with mean $a$ and variance $\sigma^2$. We also assume that $r_t, t=0,1,\dots,T-1$, are statistically independent. Investors enter the market with initial wealth $x_0$, and they wish the expected value of the terminal wealth $x_T$ will be $b$ under the policy $\boldsymbol{u} = \{u_0,u_1,\dots,u_{T-1}\}$. Briefly, investors will face the following optimization problem,
\begin{align*}
\min ~ & \Var(x_T^u), \\
\mbox{s.t. } & \E[x_T^u] = b, \\
& x_{t+1} = r_f x_t + r_t u_{t}, \quad\quad t=0,1,\dots, T-1.
\end{align*}

To solve this problem, we can first transform it into an unconstrained problem by introducing a Lagrange multiplier $w$,
\begin{align}\label{1}
\min_{\boldsymbol{u}} ~ \E(x_T^u)^2-b^2 -2w(\E[x_T^u]-b) = \min_{\boldsymbol{u}} ~ \E(x_T^u - w )^2-(w-b)^2.
\end{align}

This problem can be solved analytically and its solution $\boldsymbol{u}^* = \{u_0^{*},u_1^{*},\dots,u_{T-1}^{*}\}$ depends on $w$. Readers interested in it can refer to \cite{li2000} for a detailed derivation.

\subsection{Discrete-time exploratory mean-variance problem}
In the reinforcement learning framework, the control process $\boldsymbol{u} = \{u_t,0 \leq t < T \}$ is randomized, which represents exploration and learning, leading to a measure-valued or distributional control process whose density function is denoted by $\boldsymbol{\pi} = \{\pi_t,0 \leq t < T \}$. Thus, the dynamics of wealth become
\begin{align}\label{2}
x_{t+1}^{\pi} = r_f x_t^{\pi} + r_{t}u_{t}^{\pi}.
\end{align}

Suppose excess return $r_t$ subjects to a distribution with mean $a$ and variance $\sigma^2$. And $u_{t}^{\pi}$ is a random control process whose probability density is $\pi_{t}$. We also assume that $r_t$ is independent of $u_{t}^{\pi}$. At period $t$, the conditional first and second moment of $ r_{t}u_{t}^{\pi}$ are denoted by
\begin{align*}
& \E_{t}[r_{t}u_{t}^{\pi}] = \E_{t}[r_{t}] \E_{t}[u_{t}^{\pi}] = a\int_{\R} u \pi_{t}(u) du,\\
&\E_{t}[(r_{t}u_{t}^{\pi})^2] = \E_{t}[(r_{t})^2] \E_{t}[(u_{t}^{\pi})^2] = (a^2 + \sigma^2) \int_{\R} u^2 \pi_{t}(u) du.
\end{align*}

The randomized, distributional control process $\boldsymbol{\pi} = \{\pi_t,0 \leq t < T \}$ is to model exploration, whose overall level is captured by its accumulative entropy
\begin{align*}
\H(\boldsymbol{\pi}) := -\sum_{t=0}^{T-1} \int_{\R} \pi_t(u) \ln\pi_t(u) du.
\end{align*}

In the discrete-time market setting, the objective function of exploratory MV problem becomes
\begin{align}\label{3}
V^{\boldsymbol{\pi}}  = \min_{\boldsymbol{\pi}}~ \E \left[(x_T^{\pi} -w)^2 +\lambda \sum_{t=0}^{T-1} \int_{\R} \pi_t(u) \ln\pi_t(u) du \right] - (w-b)^2,
\end{align}
where the temperature parameter $\lambda $ measures the trade-off between exploitation and exploration in this MV problem. In the following we will derive the optimal feedback controls and the corresponding optimal value functions of the problem (\ref{3}) theoretically. We define the value function $J(t,x;w)$ under any given policy $\boldsymbol{\pi}$:
\begin{align*}
J(t,x;w)=\E \left[ (x_T^{\pi}-w)^2+\lambda \sum_{s=t}^{T-1}\int_{\R}\pi_s(u) \ln \pi_s(u) du\Big| x_t^{\pi} = x\right]-(w-b)^2.
\end{align*}

The function $J^*(t,x;w)$ is called the optimal value function of problem (\ref{3}):
\begin{align*}
J^*(t,x;w)=\min_{\pi_t,\dots,\pi_{T-1}}\E \left[ (x_T^{\pi}-w)^2+\lambda \sum_{s=t}^{T-1}\int_{\R}\pi_s(u) \ln \pi_s(u) du\Big| x_t^{\pi} = x\right]-(w-b)^2,
\end{align*}
with $J^*(T,x;w)=(x-w)^2-(w-b)^2$.

\begin{theorem}
At period t, the optimal value function is given by
\begin{align}
&\quad J^*(t,x;w)\notag \\
&=\left(\frac{\sigma^2 r_f^2}{a^2 + \sigma^2} \right)^{T-t} ( x-\rho_t w)^2 +\frac{\lambda}{2}(T-t)\ln \left( \frac{a^2 + \sigma^2}{\pi \lambda} \right) +\frac{\lambda}{2}\sum_{i=t+1}^{T}(T-i)\ln \left(  \frac{\sigma^2 r_f^2}{a^2 + \sigma^2} \right)\notag \\
& \quad-(w-b)^2,
\end{align}
where $\rho_t=( r_f^{-1})^{T-t}$. Moreover, the optimal feedback control is Gaussian, with its density function given by
\begin{align}
\pi^*(u;t,x,w)
&= \mathcal{N} \left(u \Big|-\frac{ar_f( x-\rho_tw)}{a^2 + \sigma^2}, \frac{\lambda}{2(a^2 + \sigma^2)} \left(\frac{a^2 + \sigma^2}{\sigma^2 r_f^2} \right)^{T-t-1} \right).
\end{align}
\end{theorem}

There are two important points to note after proving Theorem 1. First, the variance of the optimal Gaussian policy measuring the level of exploration is $\frac{\lambda}{2(a^2 + \sigma^2)} \left(\frac{a^2 + \sigma^2}{\sigma^2 r_f^2} \right)^{T-t-1}$ at time $t$, which means the exploration decays in time. The agent initially engages in exploration at the maximum level and reduces it gradually as time passes. Exploitation dominates exploration and becomes more important as time approaches maturity because there is a deadline $T$ at which investors' actions will be evaluated. Second, the mean of the optimal Gaussian policy is independent of the exploration weight $\lambda$, while its variance is independent of the state $x$. This fact highlights a perfect separation between exploitation and exploration, as the former is captured by the mean and the latter by the variance of the optimal Gaussian policy.

\section{Discrete-Time Algorithm}
Having proved the optimal solution of the discrete-time exploratory mean-variance problem, we next design a RL algorithm to learn the solution and output portfolio allocation strategies directly. We first need to establish a policy improvement theorem and the corresponding convergence result. Besides, we will also design a self-correcting scheme to learn the true Lagrange multiplier $w$. It is worth noting that this RL algorithm skips the estimation of model parameters, such as the mean and variance of the excess return $r_t$, which are hard to estimate accurately.

\subsection{Policy improvement and policy convergence}
Policy improvement scheme can update the current policy in the right direction to improve the value function. Thus, a policy improvement theorem will ensure the iterated value functions to be nonincreasing (in the case of a minimization problem) and converge to the optimal value function in the end. The following result provides a policy improvement theorem for the discrete-time exploratory mean-variance portfolio selection problem.

\begin{theorem}
Suppose $w$ $\in$ $\R$ is fixed and $\boldsymbol{\pi}^0$ is an arbitrarily given admissible feedback control policy, subjecting to
\begin{align*}
\pi^0_t(u;x,w)=\mathcal{N} \left(u \Big|K(x-\rho_t w),\lambda B C^{T-t-1} \right)
\end{align*}

Then we can calculate $J^{\pi^0}(t,x;w)$, where $\rho_t = r_f^{-(T-t)}$, $A=r_f^2+(a^2+\sigma^2)K^2+2r_f aK$ and $f(t) = \frac{\lambda B (a^2+\sigma^2)[1-(CA)^{T-t}]}{1-CA}-\frac{\lambda}{2}\ln(2\pi \lambda B) (T-t)-\frac{\lambda}{2}(T-t)-\frac{\lambda}{2}\ln C \sum_{i=0}^{T-t-1}i-(w-b)^2$ is a smooth function that only depends on $t$ 
\begin{align*}
&\quad J^{\pi^0}(t,x;w)\\ 
&=A^{T-t}(x-\rho_t w)^2+\frac{\lambda B (a^2+\sigma^2)[1-(CA)^{T-t}]}{1-CA}-\frac{\lambda}{2}\ln(2\pi \lambda B) (T-t)-\frac{\lambda}{2}(T-t)\\
&\quad -\frac{\lambda}{2}\ln C \sum_{i=0}^{T-t-1}i-(w-b)^2\\
&=A^{T-t}(x-\rho_t w)^2+f(t)
\end{align*}
Using the condition $\pi_t^{k+1}(u;x,w) = \arg \mathop{\min}\limits_{\pi^k_t(u)} J^{\pi^k}(t,x;w)$ to update the feedback policy and making this iteration for $k$ times, we can get $\pi_t^k(u;x,w)$ and the corresponding value function $J^{\pi^k}(t,x;w)$:
\begin{align}
\pi_t^k(u;x,w)=\mathcal{N} \left(u \Big|-\frac{a r_f (x-\rho_t w)}{a^2 + \sigma^2}, \frac{\lambda}{2(a^2 + \sigma^2)A^{T-t-k}}\left(\frac{a^2+\sigma^2}{\sigma^2 r_f^2}\right)^{k-1} \right),
\end{align}
\begin{align}
&\quad J^{\pi^k}(t,x;w)\notag \\
&=A^{T-t-k}\left(\frac{\sigma^2 r_f^2}{a^2+\sigma^2}\right)^k (x-\rho_t w)^2+\frac{\lambda}{2} k \ln\left(\frac{a^2+\sigma^2}{\pi \lambda}\right)+\frac{\lambda}{2} \sum_{i=0}^{k-1} i\ln\left(\frac{\sigma^2 r_f^2}{a^2+\sigma^2}\right) \notag \\
& \quad+\frac{\lambda \ln A}{2} k(T-t-k)+f(t+k). 
\end{align}
\end{theorem}
The policy improvement and policy convergence theorem has been proved, suggesting that there are always policies in the Gaussian family improving the value function of any given policy under the iteration condition $\pi_t^{k+1}(u;x,w) = \arg \mathop{\min}\limits_{\pi^k_t(u)} J^{\pi^k}(t,x;w)$. It turns out that, theoretically, such a updating scheme leads to the convergence of both the value functions and the policies in a finite number of iterations ($T-t$, in fact).

\subsection{Algorithm Design} 

In this section, we provide a RL algorithm, the discrete-time EMV algorithm to solve the discrete-time exploratory portfolio selection problem. This algorithm contains three procedures: policy evaluation, policy improvement and a self-correcting scheme for learning the Lagrange multiplier $w$.

For the policy evaluation, we have the following result by Bellman equation
\begin{align*}
J^{\pi}(t,x;w) = \E \left[J^{\pi}(t+1,x_{t+1};w)+\lambda \int_{\R}\pi_t(u) \ln \pi_t(u)du \Big| x_t = x \right].
\end{align*}
Rearranging this equation, we obtain
\begin{align*}
\E \left[J^{\pi}(t+1,x_{t+1};w)-J^{\pi}(t,x_t;w)+\lambda \int_{\R}\pi_t(u) \ln \pi_t(u)du \Big| x_t = x \right]=0.
\end{align*}

We define the Bellman's error
\begin{align*}
\delta_t = \hat{J}_t^{\pi}+\lambda \int_{\R}\pi_t(u) \ln \pi_t(u)du,
\end{align*}
where $\hat{J}_t^{\pi}=J^{\pi}(t+1,x_{t+1};w)-J^{\pi}(t,x_t;w)$.

The objective of the policy evaluation is to minimize the Bellman's error $\delta_t$. $J^{\theta}$ and $\pi^{\phi}$ are the parameterized value function and policy respectively, with $\theta$ and $\phi$ being the vector of parameters to be learned. We then minimize
\begin{align*}
C(\theta,\phi)&=\frac{1}{2}\E \left[\sum_{i=0}^{T-1}|\delta_i|^2 \right]\\
&=\frac{1}{2}\E \left[\sum_{i=0}^{T-1}|J^{\theta}(i+1,x_{i+1};w)-J^{\theta}(i,x_i;w)+\lambda \int_{\R}\pi_i^{\phi}(u) \ln \pi_i^{\phi}(u)du|^2 \right]\\
&=\frac{1}{2}\E \left[\sum_{i=0}^{T-1}|\hat{J}^{\theta}_i+\lambda \int_{\R}\pi_i^{\phi}(u) \ln \pi_i^{\phi}(u)du|^2 \right],
\end{align*}
where $\pi^{\phi}=\{\pi^{\phi}_t, t=0,1,\dots,T-1\}$. We collect a set of samples $D=\{(t,x_t), t=0,1,\dots,T\}$ in the following way. The initial sample is $(0,x_0)$ at $t=0$. At $t=0,1,\dots,T-1$, we sample $\pi^{\phi}_t$ to obtain an allocation $u_t$ in the risky asset, and observe the wealth $x_{t+1}$ at next time $t+1$. We can approximate $C(\theta,\phi)$ by
\begin{align*}
C(\theta,\phi)&=\frac{1}{2}\sum_{(t,x_t)\in D}\left[J^{\theta}(t+1,x_{t+1};w)-J^{\theta}(t,x_t;w)+\lambda \int_{\R}\pi_t^{\phi}(u) \ln \pi_t^{\phi}(u)du \right]^2 \\
&=\frac{1}{2}\sum_{(t,x_t)\in D}\left[\hat{J}^{\theta}(t,x_t;w)+\lambda \int_{\R}\pi_t^{\phi}(u) \ln \pi_t^{\phi}(u)du \right]^2.
\end{align*}

As suggested by the theoretical optimal value function, we consider the parameterized $J^{\theta}(t,x;w)$ by
\begin{align*}
J^{\theta}(t,x;w)=\theta_1^{T-t}(x-\rho_t w)^2+\theta_2 t^2+\theta_3 t+\theta_4,
\end{align*}
where $\rho_t=r_f^{-(T-t)}$ and $r_f$ is known.

From the policy improvement updating scheme, it follows that the mean of the policy $\pi^{\phi}_t(u)$ is $-\frac{a r_f (x-\rho_t w)}{a^2 + \sigma^2}$ and the variance of the policy is $\frac{\lambda}{2(a^2 + \sigma^2)\theta_1^{T-t-1}}$, resulting in the entropy
\begin{align*}
H(\pi^{\phi}_t)&=-\int_{\R}\pi_t^{\phi}(u) \ln \pi_t^{\phi}(u)du \\
&=\frac{1}{2} \ln \left(\frac{\pi e \lambda}{a^2+\sigma^2}\right)-\frac{\ln \theta_1}{2}(T-t-1).
\end{align*}

Equating this with the form $H(\pi^{\phi}_t)=\phi_1+\phi_2 (T-t-1)$, we can deduce
\begin{align*}
    \begin{cases}
    \phi_1 = \frac{1}{2} \ln (\frac{\pi e \lambda}{a^2+\sigma^2}), \\
    \phi_2 = -\frac{\ln \theta_1}{2}, \\
    \theta_1 = \frac{\sigma^2 r_f^2}{a^2+\sigma^2}. 
    \end{cases}
\end{align*}
The improved policy in turn becomes
\begin{align*}
&\quad \pi^{\phi}_t(u;x,w)\\
&=\mathcal{N} \left(u \Big|-\frac{a r_f (x-\rho_t w)}{a^2 + \sigma^2}, \frac{\lambda}{2(a^2 + \sigma^2)\theta_1^{T-t-1}}\right)\\
&=\mathcal{N} \left(u \Big|-\sqrt{\frac{r_f^2-e^{-2\phi_2}}{\lambda \pi}}e^{\frac{2\phi_1 -1}{2}}(x-\rho_t w),\frac{1}{2\pi}e^{2\phi_2 (T-t-1)+2\phi_1 -1} \right).
\end{align*}
Rewriting $C(\theta,\phi)$ using $H(\pi^{\phi}_t)=\phi_1+\phi_2 (T-t-1)$, we obtain
\begin{align*}
C(\theta,\phi)=\frac{1}{2}\sum_{(t,x_t)\in D}\left[\hat{J}^{\theta}(t,x_t;w)-\lambda (\phi_1+\phi_2 (T-t-1)) \right]^2.
\end{align*}
Note that $\hat{J}^{\theta}(t,x_t;w)=J^{\theta}(t+1,x_{t+1};w)-J^{\theta}(t,x_t;w)$, with $\theta_1=e^{-2\phi_2}$ in the parametrization of $J^{\theta}(t,x_t;w)$. It is now straightforward to devise the updating rules for $(\phi_1,\phi_2)^{'}$ and $(\theta_2,\theta_3)^{'}$ using stochastic gradient descent algorithms.
\begin{align}\label{7}
\frac{\partial C}{\partial \theta_2}=\sum_{(t,x_t)\in D}\left[\hat{J}^{\theta}(t,x_t;w)-\lambda (\phi_1+\phi_2 (T-t-1)) \right] ((t+1)^2-t^2),
\end{align}
\begin{align}\label{8}
\frac{\partial C}{\partial \theta_3}=\sum_{(t,x_t)\in D}\left[\hat{J}^{\theta}(t,x_t;w)-\lambda (\phi_1+\phi_2 (T-t-1)) \right], 
\end{align}
\begin{align}\label{9}
\frac{\partial C}{\partial \phi_1}=-\lambda \sum_{(t,x_t)\in D}\left[\hat{J}^{\theta}(t,x_t;w)-\lambda (\phi_1+\phi_2 (T-t-1)) \right], 
\end{align}
\begin{align}\label{10}
\frac{\partial C}{\partial \phi_2}&=\sum_{(t,x_t)\in D}\left[\hat{J}^{\theta}(t,x_t;w)-\lambda (\phi_1+\phi_2 (T-t-1)) \right]\times \Big[2(T-t)e^{-2\phi_2 (T-t)}(x_t-\rho_t w)^2 \notag\\
&\quad -2(T-t-1)e^{-2\phi_2 (T-t-1)}(x_{t+1}-\rho_{t+1} w)^2-\lambda(T-t-1) \Big].
\end{align}
Moreover, the parameter $\theta_1$ is updated with $\theta_1=e^{-2\phi_2}$, and $\theta_4$ is updated based on the terminal
condition $J^{\theta}(T,x;w)=(x-w)^2-(w-b)^2$, which yields
\begin{align}\label{11}
\theta_4=-\theta_2 T^2-\theta_3 T-(w-b)^2.
\end{align}

Finally, we provide a scheme for learning the Lagrange multiplier $w$ with $\alpha$ being the learning rate
\begin{align}
w_{n+1}=w_n-\alpha(x_T-b).
\end{align}
In implementation, we replace $x_T$ by the sample average $\frac{1}{N}\sum_j x_T^j$, where N is the sample size and $\{x_T^j\}$ are the most recent $N$ terminal wealth values obtained at the time when $w$ is to be updated.

We now summarize the pseudocode for the discrete-time EMV algorithm.

\begin{algorithm}[H]
\DontPrintSemicolon
  \SetAlgoLined
  \KwIn {Market, learning rates $\alpha, \eta_{\phi}, \eta_{\theta}$, initial wealth $x_0$, target payoff $b$, investment horizon $T$, exploration rate $\lambda$, number of iterations $M$, sample average size $N$.}
  Initialize $\theta, \phi$ and $w$ \;
  \For{$k=1$ to M}{
  Sample $\{(j,x_j^k),0 \leq j \leq T\}$ from market under $\pi^{\phi}$ \;
  \For{$i=1$ to T}{
  Obtain collected samples $D=\{(j,x_j^k),0 \leq t \leq i \}$ \;
  Update $\theta \gets \theta-\eta_{\theta} \nabla_{\theta} C(\theta,\phi)$ using $(\ref{7})$ and $(\ref{8})$\;
  Update $\theta_4$ using $(\ref{11})$ and $\theta_1 \gets e^{-2\phi_2}$\;
  Update $\phi \gets \phi-\eta_{\phi} \nabla_{\phi} C(\theta,\phi)$ using $(\ref{9})$ and $(\ref{10})$\;
  }
  Update $\pi_t^{\phi} \gets \mathcal{N} \left(u \Big|-\sqrt{\frac{r_f^2-e^{-2\phi_2}}{\lambda \pi}}e^{\frac{2\phi_1 -1}{2}}(x-\rho_t w),\frac{1}{2\pi}e^{2\phi_2 (T-t-1)+2\phi_1 -1} \right)$\;
  \eIf{k mod N == 0}{
  Update $w \gets w-\alpha\left(\frac{1}{N}\sum_{j=k-N+1}^k x_T^j -b\right)$\;
  }{
  $w \gets w$\;
  }
  }
  \caption{Discrete-Time Mean-Variance Portfolio Selection}
\end{algorithm}

\section{Simulation Study}
In this section, we carry out simulations to compare the performance of our RL algorithm in the discrete-time setting with the EMV algorithm established by \cite{zhou2020mv}. We conduct comparisons in the market setting where the excess return of the risky asset subjects to the skewed $T$ distribution rather than normal distribution. The reason why we choose the skewed $T$ distribution is that it can better reflect skewness and tail risk. In the following, we first describe the RL algorithm in \cite{zhou2020mv} briefly.

\subsection{EMV algorithm in \cite{zhou2020mv}}
\cite{zhou2020mv} provided a RL algorithm (EMV algorithm) to solve the continuous-time mean-variance portfolio selection problem. As for the market setting, they assumed that the price process of the risky asset is a geometric Brownian motion with $S_0 = s_0 > 0$ being the initial price at $t = 0$, and $\mu, \sigma$ being the mean and volatility parameters respectively. Thus, the Sharpe ratio of the risky asset is defined by $\rho = \frac{\mu-r}{\sigma}$, with $r$ being the constant interest rate of the riskless asset. Different from the classical continuous-time MV problem, the exploratory dynamics of the wealth is changed to
\begin{align*}
dX_t^{\pi} = \rho \sigma \mu_t dt + \sigma \sqrt{\mu_t^2+\sigma_t^2} dW_t,
\end{align*}
where $\mu_t$ and $\sigma_t^2$ are the mean and variance of the control process with distribution $\pi_t$ respectively. They apply the classical Bellman's principle of optimality and the Hamilton-Jacobi-Bellman (HJB) equation to derive the optimal feedback control which is Gaussian and the optimal value function. A policy improvement theorem has also been proved, which is a crucial prerequisite for an interpretable RL algorithm. Having laid the theoretical foundation, authors presented EMV algorithm which consists of three procedures: the policy evaluation, the policy improvement and a self-correcting scheme for learning the Lagrange multiplier $w$ based on stochastic approximation. 

In implementation, they discretize the time period $[0,T]$ and $\Delta t$ is the discretization step for the learning algorithm. Taking advantage of the explicit parametric expressions of the value function and control policy, they consider the parameterized $V^{\theta}$ and $\boldsymbol{\pi}^{\phi}$. Based on the policy improvement updating scheme and policy improvement theorem, this RL algorithm will finally figure out the optimal feedback controls and corresponding optimal value functions. They actually skip the process of using historical data to estimate the market parameters $\mu$ and $\sigma$ and just need to interact with the environment to update the parameterized $V^{\theta}$ and $\boldsymbol{\pi}^{\phi}$. And they prove that their method outperforms two other methods, MLE and DDPG, both in simulation and empirical studies.

\subsection{The stationary market case}
\begin{figure}[H]
    \begin{minipage}{\linewidth}
    \centering
    \includegraphics{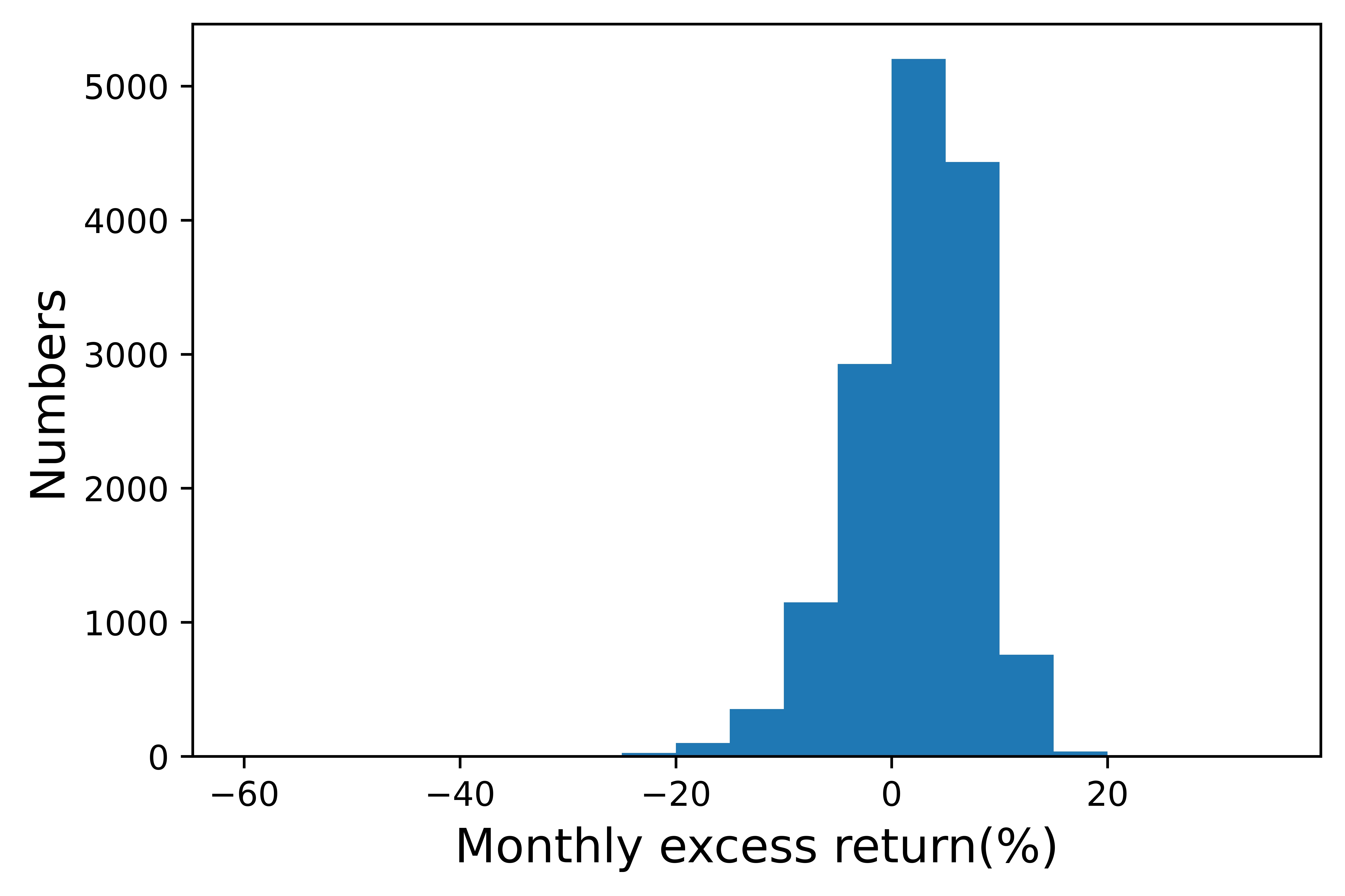}
    \caption{The histogram of a risky asset’s monthly excess return subjecting to the skewed t distribution}
    \label{Fig1}
    \end{minipage}
\end{figure}

\cite{Theodossiou1998} concluded that the skewed generalized
$t$ distribution provided an excellent fit to the empirical distribution of the financial data, such as U.S. stock market returns. We perform numerical simulation in a stationary market environment, where the excess return of the risky asset is simulated according to the skewed $t$ distribution. Figure \ref{Fig1} presents the histogram of a risky asset’s monthly excess returns subjecting to skewed t distribution with annualized excess return $a = 30\%$, volatility $\sigma = 20\%$ , degree of freedom being $10$ and skewness parameter being $-1.5$, which can better reflect the tail risk, especially extreme negative return. We take $T = 3$ and $\Delta t = 1$, indicating that the MV problem is considered over a 3-month period, with monthly rebalancing. We choose 30\% as the annualized excess return of the risky asset and volatility will be taken from the set $\sigma \in \{10\%, 20\%, 30\% \}$. The annualized interest rate $r_f$ is taken to be 2\%, which is known and fixed. We consider the discrete-time exploratory MV problem with a 10\% target return on the terminal wealth over a 3-month period, starting from a normalized initial wealth $x_0 = 1$ and $b = 1.1$. Besides, we take the number of iterations $M = 15000$ and take the sample average size $N = 10$ to update the Lagrange multiplier $w$. The exploration rate $\lambda = 2$ measures the trade-off between exploitation and exploration in this MV problem. The learning rates $\alpha = 0.05$ and $\eta_{\phi} = \eta_{\theta} = 0.0005$ respectively. All these parameters will be fixed for all the simulations in this section.

To compare the performances of two algorithms under each market scenario, we present the return and standard deviation of the last 2000 values of the terminal wealth, and the corresponding Sharpe ratio. In other words, we divide the 15000 episodes into two parts: the first 13000 episodes as the training set and the last 2000 episodes as the testing set. We evaluate the performances of two methods through the their out-of-sample performances.

\begin{table}[H] 
\setlength{\abovecaptionskip}{0mm}
\setlength{\belowcaptionskip}{3mm}
\centering
\caption{Comparison of the return, standard deviation and Sharpe ratio for two algorithms in simulations}
\begin{tabular}{ccc}   
\hline   \textbf{Market Setting} & \textbf{Discrete-Time Algorithm} & \textbf{EMV Algorithm} \\ 
\hline   $a = 30\%, \sigma = 10\%$ & 10.0\%;~8.5\%;~1.18 & 9.1\%;~8.0\%;~1.14  \\ 
$a = 30\%, \sigma = 20\%$ & 9.9\%;~14.4\%;~0.68 & 9.1\%;~13.8\%;~0.66  \\  
$a = 30\%, \sigma = 30\%$ & 10.4\%;~20.0\%;~0.52 & 9.3\%;~18.8\%;~0.49\\ 
\hline
\end{tabular}    
\end{table}

From Table 1, in all the three simulation experiments, it is clear that our discrete-time algorithm can achieve better out-of-sample performance in mean return and Sharpe ratio taking account into 10\% target return set before compared with the EMV algorithm in \cite{zhou2020mv}.

\begin{figure}[H]
    \begin{minipage}{\linewidth}
    \centering
    \includegraphics{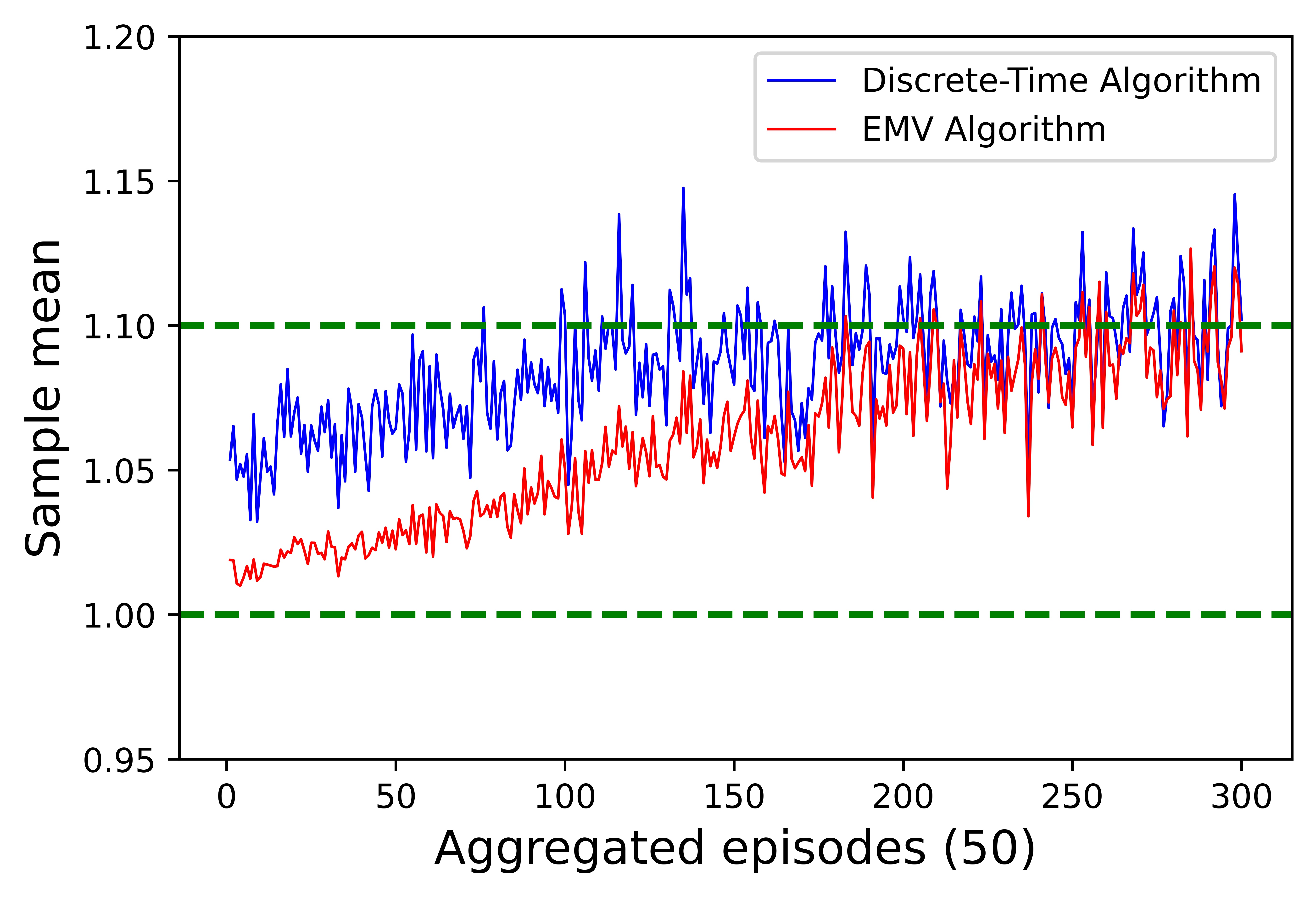}
    \caption{Learning curves of sample means of terminal wealth (over every 50 iterations) for two algorithms (a=30\%, $\sigma$=10\%)}
    \label{Fig2}
    \end{minipage}
\end{figure}

\begin{figure}[H]
    \begin{minipage}{\linewidth}
    \centering
    \includegraphics{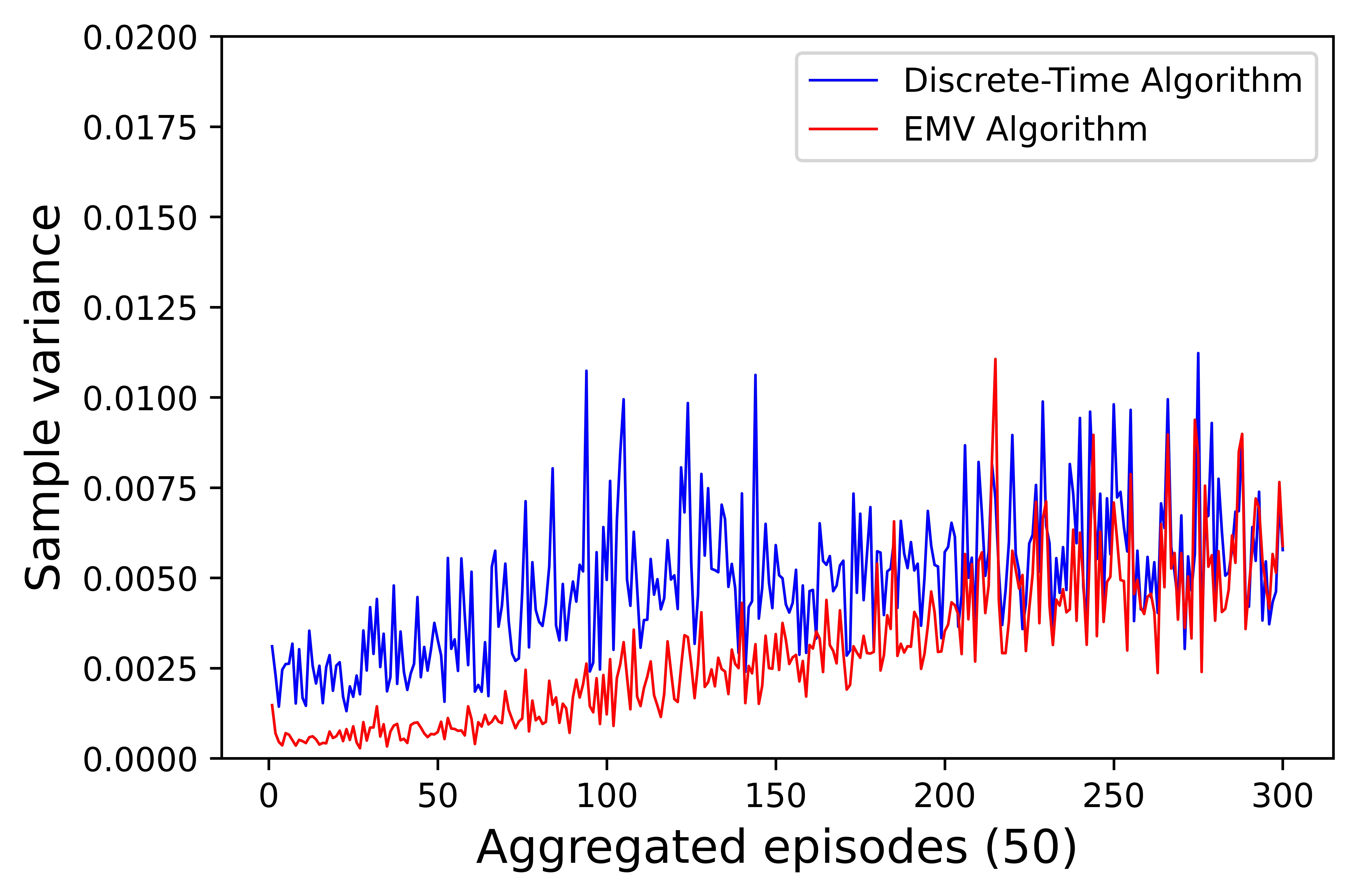}
    \caption{Learning curves of sample variances of terminal wealth (over every 50 iterations) for two algorithms (a=30\%, $\sigma$=10\%)}
    \label{Fig3}
    \end{minipage}
\end{figure}
We also present the changes of the sample mean and sample variance of every nonoverlapping 50 terminal wealth values ($a = 30\%, \sigma = 10\%$). From Figure \ref{Fig2} and Figure \ref{Fig3}, it is clear that our algorithm converges relatively faster than EMV algorithm in \cite{zhou2020mv}, achieving relatively stable performance in the early phase of the learning process. This is also consistent with the policy convergence result proved before.

\section{Empirical Study}
\begin{figure}[H]
    \begin{minipage}{\linewidth}
    \centering
    \includegraphics{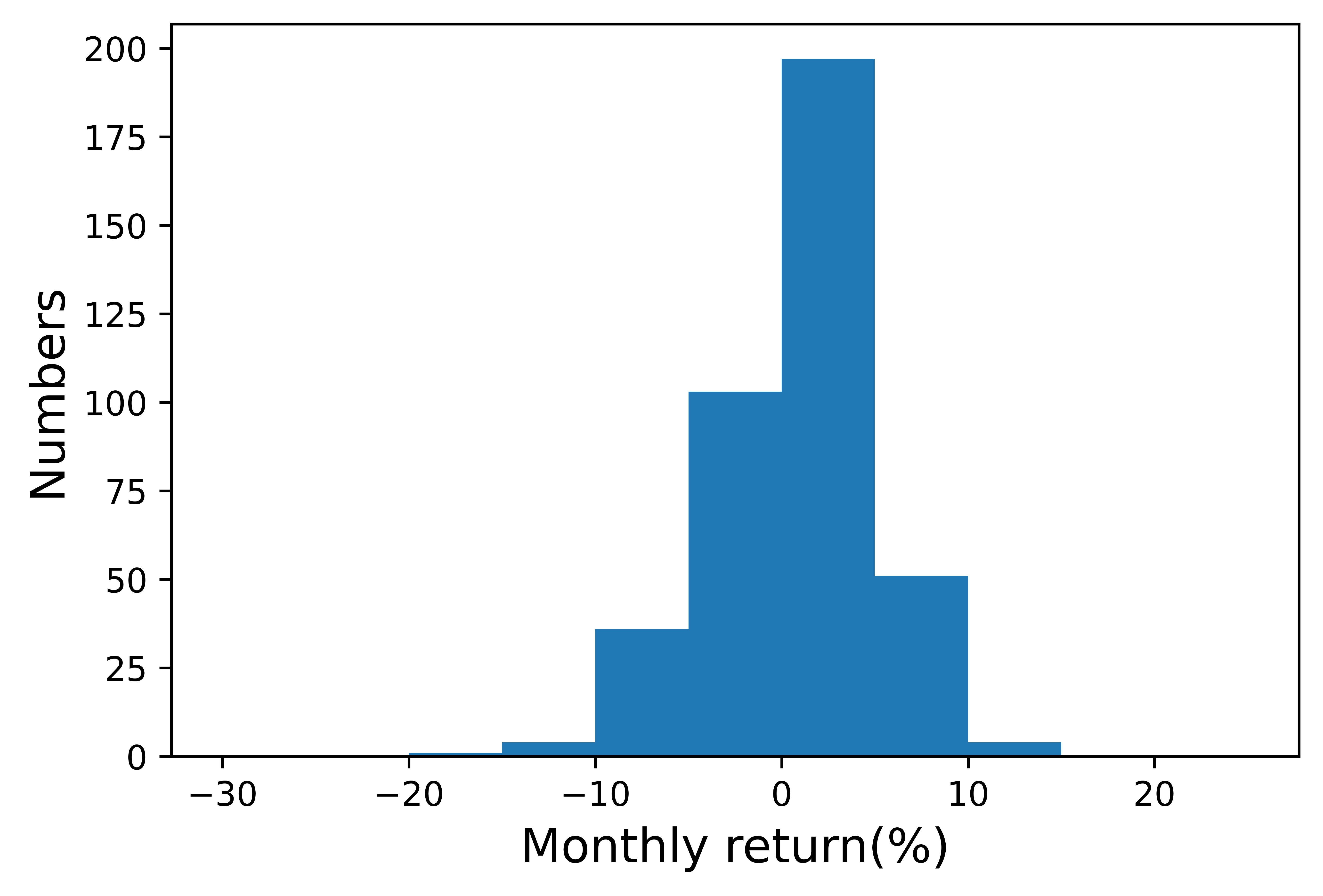}
    \caption{The histogram of S$\&$P 500 index's monthly return during the period of 1990-2022}
    \label{Fig4}
    \end{minipage}
\end{figure}

In this section, we examine the efficiency of our RL algorithm on real-world financial data and compare its performance against the EMV algorithm in \cite{zhou2020mv}. We conduct our empirical analysis in the one-dimensional case, which means that we just consider a market containing one risky asset and one risk-free asset. We choose S\&P 500 index as the risky asset and set the annualized return of the risk-free asset to be 2\%, which is fixed. We have plotted the histogram of S$\&$P 500's monthly return during the period of 1990-2022. From Figure \ref{Fig4}, it is clear that monthly returns of S$\&$P 500 mainly concentrate in the interval ranging from 0 to 5\%. Besides, there is an obvious characteristic--fat tail--needing to be highlighted. Extremely positive and negative monthly returns of a risky financial asset are rare, however, they do happen in the era of boom or recession.

In the empirical study, we investigate the MV problem over a 3-month period with monthly rebalancing. And we consider this problem starting from a normalized initial wealth $x_0 =1$ and set the 3-month target to be $b = 1.05$ (corresponding to a 21.55\% annualized return). Learning rates, $\alpha = 0.05$ and $\eta_{\phi} = \eta_{\theta} = 0.0005$, are as the same with the simulation study.
To check the robustness of the performance comparison of the two algorithms, we repeat the same test for 10 rolling 10-year horizons. The training set contains the 10-year monthly data immediately before each testing period. After updating the parameters of the value functions and policies in the training set, we then evaluate the out-of-sample performance of two algorithms in the testing set. From Table 2, it is clear that with Sharpe ratios being nearly equal, our discrete-time algorithm can achieve higher mean return compared with the EMV algorithm. Besides, we will also provide the performance comparison under different return targets ($b = 1.03$ and $b = 1.07$).
\begin{table}[H] 
\setlength{\abovecaptionskip}{0mm}
\setlength{\belowcaptionskip}{3mm}
\centering
\caption{Comparison of the return, standard deviation and Sharpe ratio for two algorithms in the empirical study when using S\&P500 index data ($b = 1.05$)}
\begin{tabular}{ccc}   
\hline   \textbf{Testing Period} & \textbf{Discrete-Time Algorithm} & \textbf{EMV Algorithm} \\ 
\hline 2004-2013 & 3.12\%;~24.17\%;~0.13 & 2.93\%;~22.08\%;~0.13 \\ 
2005-2014 & 2.76\%;~17.26\%;~0.16 & 2.71\%;~16.70\%;~0.16  \\  
2006-2015 & 3.23\%;~23.10\%;~0.14 & 3.07\%;~21.59\%;~0.14\\ 
2007-2016 & 3.25\%;~27.22\%;~0.12 & 3.07\%;~25.13\%;~0.12  \\  
2008-2017 & 6.07\%;~33.86\%;~0.18 & 5.57\%;~30.55\%;~0.18\\ 
2009-2018 & 6.60\%;~13.25\%;~0.50 & 5.31\%;~10.47\%;~0.51  \\  
2010-2019 & 10.98\%;~23.09\%;~0.48 & 7.57\%;~15.68\%;~0.48\\ 
2011-2020 & 7.24\%;~16.64\%;~0.44 & 5.56\%;~12.55\%;~0.44  \\  
2012-2021 & 12.89\%;~22.72\%;~0.57 & 10.96\%;~19.27\%;~0.57\\ 
2013-2022 & 8.45\%;~20.43\%;~0.41 & 7.64\%;~18.46\%;~0.41\\ 
\hline
\end{tabular} 
\end{table}

\begin{table}[H] 
\setlength{\abovecaptionskip}{0mm}
\setlength{\belowcaptionskip}{3mm}
\centering
\caption{Comparison of the return, standard deviation and Sharpe ratio for two algorithms in the empirical study when using S\&P500 index data ($b = 1.03$)}
\begin{tabular}{ccc}   
\hline   \textbf{Testing Period} & \textbf{Discrete-Time Algorithm} & \textbf{EMV Algorithm} \\ 
\hline 2004-2013 & 1.97\%;~13.01\%;~0.15 & 1.89\%;~12.25\%;~0.15 \\ 
2005-2014 & 1.82\%;~9.79\%;~0.19 & 1.77\%;~9.40\%;~0.19  \\  
2006-2015 & 2.01\%;~12.43\%;~0.16 & 1.91\%;~11.56\%;~0.17\\ 
2007-2016 & 2.04\%;~14.73\%;~0.14 & 1.94\%;~13.72\%;~0.14  \\  
2008-2017 & 3.57\%;~18.15\%;~0.20 & 3.27\%;~16.32\%;~0.20\\ 
2009-2018 & 4.18\%;~8.04\%;~0.52 & 3.76\%;~7.14\%;~0.53  \\  
2010-2019 & 8.96\%;~18.75\%;~0.48 & 7.04\%;~14.55\%;~0.48\\ 
2011-2020 & 4.57\%;~10.14\%;~0.45 & 3.79\%;~8.22\%;~0.46  \\  
2012-2021 & 8.85\%;~15.44\%;~0.57 & 8.16\%;~14.19\%;~0.58\\ 
2013-2022 & 4.78\%;~11.18\%;~0.43 & 4.44\%;~10.33\%;~0.43\\ 
\hline
\end{tabular} 
\end{table}

\begin{table}[H] 
\setlength{\abovecaptionskip}{0mm}
\setlength{\belowcaptionskip}{3mm}
\centering
\caption{Comparison of the return, standard deviation and Sharpe ratio for two algorithms in the empirical study when using S\&P500 index data ($b = 1.07$)}
\begin{tabular}{ccc}   
\hline   \textbf{Testing Period} & \textbf{Discrete-Time Algorithm} & \textbf{EMV Algorithm} \\ 
\hline 2004-2013 & 3.95\%;~32.93\%;~0.12 & 3.92\%;~31.95\%;~0.12 \\ 
2005-2014 & 3.65\%;~24.74\%;~0.15 & 3.51\%;~23.15\%;~0.15  \\  
2006-2015 & 4.26\%;~32.70\%;~0.13 & 4.03\%;~30.14\%;~0.13\\ 
2007-2016 & 4.28\%;~38.51\%;~0.11 & 4.05\%;~35.45\%;~0.11  \\  
2008-2017 & 8.08\%;~46.95\%;~0.17 & 7.19\%;~40.76\%;~0.18\\ 
2009-2018 & 8.42\%;~17.09\%;~0.49 & 6.47\%;~12.94\%;~0.50  \\  
2010-2019 & 12.63\%;~26.59\%;~0.47 & 8.78\%;~18.30\%;~0.48\\ 
2011-2020 & 9.63\%;~22.39\%;~0.43 & 6.82\%;~15.60\%;~0.44  \\  
2012-2021 & 16.91\%;~29.88\%;~0.57 & 14.07\%;~24.85\%;~0.57\\ 
2013-2022 & 11.65\%;~28.25\%;~0.41 & 10.15\%;~24.72\%;~0.41\\ 
\hline
\end{tabular} 
\end{table}

\section{Conclusion}
In this paper, we have developed a RL framework for the discrete-time MV portfolio selection problem. By transforming the MV portfolio selection into an exploration/learning versus exploitation/optimization problem, we are able to skip the estimation of the unknown model parameters which is difficult in investment practice. The exploration part is explicitly captured by the entropy-regularized objective function of the new optimization problem. We derived the optimal value function and feedback control policy distribution which is Gaussian with a time-decaying variance.

Having derived the expression of the optimal policy and value function, we also proved a policy improvement and convergence result, which yields an updating scheme for the control policy that improves the value function in each iteration. These theoretical foundations allow us to design an effective RL algorithm to find the optimal solutions of this MV problem. The advantage of our algorithm against the EMV algorithm in \cite{zhou2020mv} has been demonstrated by numerical simulations with stationary market environments, as well as by an empirical study in the S\&P 500 index data.

Financial markets are usually unstable and hard to build some models to describe accurately, so what we know about them are approximations about reality based on our models. A RL framework to solve the MV portfolio selection problem is based on some basic assumptions about the markets. However, there is vast disagreement among finance academicians on how to best model returns. If one model deviates far from the true market, model mispecification and
estimation error will result in the policies that can't bring great investment performance in the real-world practice. The discrete-time model we proposed in this paper is based on relatively weaker and more general assumptions about the risky asset's return distribution, and it exhibits better applicability compared with the continuous-time model in \cite{zhou2020mv}.

The fact that the agent seeks to optimize future rewards means that reinforcement learning may be a valuable tool for portfolio choice, because reinforcement learning broadly consists of methods for
solving a sequential decision making problem under uncertainty. In a RL framework, an agent takes an action to maximize cumulative reward function and the state of the system is influenced by the agent's action. Asset managers and banks are often the most active participants in financial markets and their tendency to trade large volumes means that their portfolio decisions will have price impact and thus alter the state of the system. Applying a RL framework to the portfolio choice problem is still left for further thoroughly investigations.

\newpage
\appendix
\section{Proof of Theorem 1}
To solve the optimization problem (\ref{3}), we obtain the optimal value function at period $t$ according to the Bellman's principle of optimality
\begin{align}\label{4}
J^*(t,x;w) = \min_{\pi_t}\E \left[J^*(t+1,x^{\pi}_{t+1};w) +\lambda \int_{\R} \pi_t(u) \ln \pi_t(u) du \Big| x_t^{\pi}=x\right].
\end{align}

Now, we first consider MV problem from period $T$ to $T-1$. The value function at period $T-1$ can be calculated from the following equation
\begin{align*}
J(T-1,x;w) =  \E \left[(x_T^{\pi} - w )^2+\lambda \int_{\R} \pi_{T-1}(u) \ln\pi_{T-1}(u) du \Big| x_{T-1}^{\pi}=x\right]-(w-b)^2.
\end{align*}
Substituting the wealth updating process (\ref{2}) into this equation, we get
\begin{align*}
&\quad J(T-1,x;w) \\
& =  \E \left[( r_f x_{T-1}^{\pi}  + r_{T-1}u_{T-1}^{\pi} - w )^2+\lambda \int_{\R} \pi_{T-1}(u) \ln \pi_{T-1}(u) du \Big| x_{T-1}^{\pi}=x\right]-(w-b)^2\\
& = \E \Big[( r_f x_{T-1}^{\pi})^2  + (r_{T-1}u_{T-1}^{\pi})^2 + w^2 + 2 r_f x_{T-1}^{\pi}r_{T-1}u_{T-1}^{\pi} -2w r_f x_{T-1}^{\pi}\\
& \quad -2w r_{T-1}u_{T-1}^{\pi} \Big| x_{T-1}^{\pi}=x \Big]+ \lambda \int_{\R} \pi_{T-1}(u) \ln\pi_{T-1}(u) du-(w-b)^2\\
&=( r_f x)^2 + (a^2 + \sigma^2) \int_{\R} u^2 \pi_{T-1}(u) du + w^2 + 2r_f x a\int_{\R} u \pi_{T-1}(u) du -2w r_f x \\
&\quad -2w a\int_{\R} u \pi_{T-1}(u) du + \lambda \int_{\R} \pi_{T-1}(u) \ln\pi_{T-1}(u) du-(w-b)^2\\
&= \int_{\R} \left((a^2 + \sigma^2)  u^2 + (2r_f x-2w) a u + \lambda  \ln\pi_{T-1}(u)\right) \pi_{T-1}(u) du + (r_f x)^2 + w^2\\
&\quad -2w r_f x-(w-b)^2.
\end{align*}

Our goal is to minimize the value function at period $T-1$ about $\pi_{T-1}(u)$. We let $\frac{\partial J(T-1,x;w)}{ \partial \pi_{T-1}(u) } = 0$, and thus obtain the following result
\begin{align*}
(a^2 + \sigma^2)  u^2 + (2r_f x-2w) a u + \lambda  \ln\pi_{T-1}(u) + \lambda =0.
\end{align*}
Applying the usual verification technique and using the fact that
$\pi \in \mathcal{P}(\R)$ if and only if
\begin{align*}
\int_{\R} \pi(u) du =1 \quad \mbox{ and } \quad \pi(u) \geq 0 ~ \mbox{ a.e on } \R, 
\end{align*}
we can solve the optimization problem and obtain a feedback control whose density function is given by
\begin{align*}
\pi^*(u;T-1,x,w) &= \frac{\exp \{-\frac{1}{\lambda} \left( (a^2 + \sigma^2)  u^2 + (2r_f x-2w) a u \right) \}}{\int_{\R}\exp \{-\frac{1}{\lambda} \left((a^2 + \sigma^2)  u^2 + (2r_f x-2w) a u \right)\}du},\\
&= \mathcal{N} \left(u \Big|-\frac{a(r_f x-w)}{a^2 + \sigma^2}, \frac{\lambda}{2(a^2 + \sigma^2)}\right).
\end{align*}

We notice that
\begin{align*}
& \E[u|\pi^*_{T-1}] = -\frac{a(r_f x-w)}{a^2 + \sigma^2} ,\\
& \E[u^2|\pi^*_{T-1}] = \left(\frac{a(r_f x-w)}{a^2 + \sigma^2}\right) ^2 + \frac{\lambda}{2(a^2 + \sigma^2)}.
\end{align*}
Substituting $\pi^*(u;T-1,x,w)$ into $J(T-1,x;w)$ leads to 
\begin{align*}
&\quad J^*(T-1,x;w) \\
&= \int_{\R} \left((a^2 + \sigma^2)  u^2 + (2r_f x-2w) a u + \lambda  \ln\pi^*_{T-1}(u)\right) \pi^*_{T-1}(u) du + ( r_f x)^2 + w^2 \\
& \quad -2w r_f x-(w-b)^2\\
&=(a^2 + \sigma^2)  \E[u^2|\pi^*_{T-1}] + (2r_f x-2w) a \E[u|\pi^*_{T-1}]  + \int_{\R} \lambda  \ln\pi^*_{T-1}(u)\pi^*_{T-1}(u) du \\
&\quad + ( r_f x-w)^2-(w-b)^2.
\end{align*}
Then we get 
\begin{align*}
& \quad \int_{\R}  \ln\pi^*_{T-1}(u)\pi^*_{T-1}(u) du \\
&= \int_{\R} \ln \left( \sqrt{\frac{2(a^2 + \sigma^2)}{2\pi\lambda}}\exp \left\{ -\frac{1}{\lambda} (a^2 + \sigma^2)(u + \frac{a(r_f x-w)}{a^2 + \sigma^2})^2 \right\}\right) \pi_{T-1}^*(u) du \\
&= \int_{\R} \left(\frac{1}{2}\ln \left( \frac{a^2 + \sigma^2}{\pi\lambda} \right)  -\frac{1}{\lambda} (a^2 + \sigma^2)(u + \frac{a(r_f x-w)}{a^2 + \sigma^2})^2 \right)\pi_{T-1}^*(u) du \\
&=\frac{1}{2}\ln \left( \frac{a^2 + \sigma^2}{\pi\lambda} \right)  -\frac{1}{\lambda} (a^2 + \sigma^2) \int_{\R} \left(u^2 + 2u\frac{a(r_f x-w)}{a^2 + \sigma^2}+ (\frac{a(r_f x-w)}{a^2 + \sigma^2})^2 \right)\pi_{T-1}^*(u) du\\
&=\frac{1}{2}\ln \left( \frac{a^2 + \sigma^2}{\pi\lambda} \right)  -\frac{1}{\lambda} (a^2 + \sigma^2) \left( \E[u^2|\pi^*_{T-1}] \!+\!2\frac{a(r_f x-w)}{a^2 + \sigma^2}\E[u|\pi^*_{T-1}] \!+\!(\frac{a(r_f x-w)}{a^2 + \sigma^2})^2\right)\\
&=\frac{1}{2}\ln \left( \frac{a^2 + \sigma^2}{\pi\lambda } \right)  -\frac{1}{\lambda}  \left( (a^2 + \sigma^2) \E[u^2|\pi^*_{T-1}] \!+\!2a(r_f x-w) \E[u|\pi^*_{T-1}] \!+ \!\frac{a^2(r_f x-w)^2}{a^2 + \sigma^2}\right).
\end{align*}
Thus, we can get %$J^*(T-1,x;w)$
\begin{align*}
&\quad J^*(T-1,x;w) \\
&=(a^2 + \sigma^2)  \E[u^2|\pi^*_{T-1}] \!+\!(2r_f x-2w) a \E[u|\pi^*_{T-1}] \!+\!\frac{\lambda}{2} \ln \left( \frac{a^2 + \sigma^2}{\pi\lambda} \right)\! -\! (a^2 + \sigma^2) \E[u^2|\pi^*_{T-1}] \\
& \quad -2a(r_f x-w) \E[u|\pi^*_{T-1}]  - \frac{a^2(r_f x-w)^2}{a^2 + \sigma^2} + ( r_f x -w)^2-(w-b)^2\\
&= \frac{\lambda}{2} \ln \left( \frac{a^2 + \sigma^2}{\pi\lambda} \right)  - \frac{a^2(r_f x-w)^2}{a^2 + \sigma^2} + ( r_f x -w)^2-(w-b)^2\\
&= \frac{\sigma^2}{a^2 + \sigma^2} (r_f x-w)^2 + \frac{\lambda}{2} \ln \left( \frac{a^2 + \sigma^2}{\pi\lambda} \right)-(w-b)^2.
\end{align*}

Next, we consider the MV problem from period $T-1$ to $T-2$. The value function at period $T-2$ can also be calculated from the following equation
\begin{align*}
&\quad J(T-2,x;w)\\
&=\E \left[J^*(T-1,x_{T-1}^{\pi};w)+\lambda\int_{\R} \pi_{T-2}(u) \ln\pi_{T-2}(u) du \Big| x_{T-2}^{\pi}=x\right] \\
&=\E \left[\frac{\sigma^2}{a^2 + \sigma^2} (r_f x_{T-1}^{\pi}-w)^2 + \frac{\lambda}{2} \ln \left( \frac{a^2 + \sigma^2}{\pi\lambda} \right)+\lambda \int_{\R} \pi_{T-2}(u) \ln\pi_{T-2}(u) du \Big| x_{T-2}^{\pi}=x\right]\\
&\quad -(w-b)^2.
\end{align*}
Substituting the wealth updating process (\ref{2}) into this equation yields 
\begin{align*}
&\quad J(T-2,x;w) \\
& =\E \Big[\frac{\sigma^2 r_f^2}{a^2 + \sigma^2} ( r_f x_{T-2}^{\pi} + r_{T-2}u_{T-2}^{\pi}- r_f^{-1}w)^2 + \lambda \int_{\R} \pi_{T-2}(u) \ln\pi_{T-2}(u) du \\
& \quad +\frac{\lambda}{2} \ln \left( \frac{a^2 + \sigma^2}{\pi\lambda} \right) \Big| x_{T-2}^{\pi}=x\Big]-(w-b)^2\\
& = \frac{\sigma^2 r_f^2}{a^2 + \sigma^2}\mathbb{E}\Big[( r_f x)^2 + (r_{T-2}u_{T-2}^{\pi})^2 + (r_f^{-1}w)^2 + 2 r_f x r_{T-2}u_{T-2}^{\pi} -2r_f^{-1}w r_f x \\
& \quad -2r_f^{-1}w r_{T-2}u_{T-2}^{\pi}  \Big]+ \lambda \int_{\R} \pi_{T-2}(u) \ln\pi_{T-2}(u) du + \frac{\lambda}{2} \ln \left( \frac{a^2 + \sigma^2}{\pi\lambda} \right)-(w-b)^2\\
&= \frac{\sigma^2 r_f^2}{a^2 + \sigma^2} \Big(( r_f x)^2 + (a^2 + \sigma^2) \int_{\R} u^2 \pi_{T-2}(u) du + (r_f^{-1}w)^2 + 2r_f x a\int_{\R} u \pi_{T-2}(u) du  \\
& \quad -2r_f^{-1}w r_f x- 2r_f^{-1}w a\int_{\R} u \pi_{T-2}(u) du \Big)+ \lambda \int_{\R} \pi_{T-2}(u) \ln\pi_{T-2}(u) du \\
&\quad + \frac{\lambda}{2} \ln \left( \frac{a^2 + \sigma^2}{\pi\lambda} \right)-(w-b)^2\\
&= \int_{\R} \left(\frac{\sigma^2 r_f^2}{a^2 + \sigma^2}  \left((a^2 + \sigma^2)  u^2 + (2r_f x-2r_f^{-1}w) a u \right)+ \lambda  \ln\pi_{T-2}(u)\right) \pi_{T-2}(u) du \\
& \quad + \frac{\sigma^2 r_f^2}{a^2 + \sigma^2}\left(( r_f x)^2 + (r_f^{-1}w)^2 -2r_f^{-1}w r_f x \right) + \frac{\lambda}{2} \ln \left( \frac{a^2 + \sigma^2}{\pi\lambda} \right)-(w-b)^2\\
&= \int_{\R} \left(  \frac{\sigma^2 r_f^2}{a^2 + \sigma^2} \left((a^2 + \sigma^2)  u^2 + (2r_f x-2r_f^{-1}w) a u \right)+ \lambda  \ln\pi_{T-2}(u)\right) \pi_{T-2}(u) du \\
& \quad + \frac{\sigma^2 r_f^2}{a^2 + \sigma^2}\left( r_f x-r_f^{-1}w \right)^2 +\frac{\lambda}{2}\ln \left( \frac{a^2 + \sigma^2}{\pi\lambda} \right)-(w-b)^2.
\end{align*}

To minimize the value function at period $T-2$ about $\pi_{T-2}(u)$, we let $\frac{\partial J(T-2,x;w)}{ \partial \pi_{T-2}(u) } = 0 $ and get
\begin{align*}
\frac{\sigma^2 r_f^2}{a^2 + \sigma^2}\left((a^2 + \sigma^2)  u^2 + (2r_f x-2r_f^{-1}w) a u \right)+ \lambda  \ln\pi_{T-2}(u) + \lambda =0.
\end{align*}
Applying the usual verification technique and using the fact that
$\pi \in \mathcal{P}(\R)$ if and only if
\begin{align*}
\int_{\R} \pi(u) du =1 \quad \mbox{ and } \quad \pi(u) \geq 0 \mbox{ a.e on } \R,
\end{align*}
we can obtain a feedback control whose density function is given by
\begin{align*}
\pi^*(u;T-2,x,w) &= \frac{\exp \{-\frac{1}{\lambda} \frac{\sigma^2 r_f^2}{a^2 + \sigma^2} \left( (a^2 + \sigma^2)  u^2 + (2r_f x-2r_f^{-1}w) a u \right) \}}{\int_{\R}\exp \{-\frac{1}{\lambda} \frac{\sigma^2 r_f^2}{a^2 + \sigma^2} \left( (a^2 + \sigma^2)  u^2 + (2r_f x-2r_f^{-1}w) a u \right) \}du}\\
&= \mathcal{N} \left(u \Big|-\frac{a(r_f x-r_f^{-1}w)}{a^2 + \sigma^2}, \frac{\lambda}{2(a^2 + \sigma^2)} \frac{a^2 + \sigma^2}{\sigma^2 r_f^2}\right).
\end{align*}

It is worth noting that
\begin{align*}
& \E[u|\pi^*_{T-2}] = -\frac{a(r_f x-r_f^{-1}w)}{a^2 + \sigma^2} ,\\
& \E[u^2|\pi^*_{T-2}] = \left(\frac{a(r_f x-r_f^{-1}w)}{a^2 + \sigma^2}\right) ^2 + \frac{\lambda}{2(a^2 + \sigma^2)}\frac{a^2 + \sigma^2}{\sigma^2 r_f^2}.
\end{align*}
Substituting $\pi^*(u;T-2,x,w)$ into $J(T-2,x;w)$ brings about 
\begin{align*}
&\quad J^*(T-2,x;w) \\
&= \int_{\R} \left(\frac{\sigma^2 r_f^2}{a^2 + \sigma^2}  \left((a^2 + \sigma^2)  u^2 + (2r_f x-2r_f^{-1}w) a u \right)+ \lambda  \ln\pi_{T-2}^*(u)\right) \pi_{T-2}^*(u) du \\
& \quad +\frac{\sigma^2 r_f^2}{a^2 + \sigma^2}\left( r_f x-r_f^{-1}w \right)^2 + \frac{\lambda}{2} \ln \left( \frac{a^2 + \sigma^2}{\pi\lambda} \right)-(w-b)^2\\
&=\frac{\sigma^2 r_f^2}{a^2 + \sigma^2} \left((a^2 + \sigma^2)  \E[u^2|\pi^*_{T-2}] + (2r_f x-2r_f^{-1}w) a \E[u|\pi^*_{T-2}] \right)\\
& \quad +\lambda \int_{\R} \ln\pi^*_{T-2}(u)\pi^*_{T-2}(u) du  + \frac{\sigma^2 r_f^2}{a^2 + \sigma^2}\left( r_f x-r_f^{-1}w \right)^2 +\frac{\lambda}{2} \ln \left( \frac{a^2 + \sigma^2}{\pi \lambda}\right)-(w-b)^2.
\end{align*}
Then we obtain 
\begin{align*}
& \quad \int_{\R}  \ln\pi^*_{T-2}(u)\pi^*_{T-2}(u) du \\
&= \int_{\R} \left(\frac{1}{2}\ln \left( \frac{a^2 + \sigma^2}{\pi \lambda} \frac{\sigma^2 r_f^2}{a^2 + \sigma^2}\right)  -\frac{1}{\lambda} (a^2 + \sigma^2)\frac{\sigma^2 r_f^2}{a^2 + \sigma^2}(u + \frac{a(r_f x-r_f^{-1}w)}{a^2 + \sigma^2})^2 \right)\pi_{T-2}^*(u) du \\
&= -\frac{1}{\lambda} (a^2 + \sigma^2) \frac{\sigma^2 r_f^2}{a^2 + \sigma^2}\int_{\R} \left(u^2 + 2u\frac{a(r_f x-r_f^{-1}w)}{a^2 + \sigma^2}+ (\frac{a(r_f x-r_f^{-1}w)}{a^2 + \sigma^2})^2 \right)\pi_{T-2}^*(u) du\\
&\quad+\frac{1}{2}\ln \left( \frac{a^2 + \sigma^2}{\pi \lambda }\frac{\sigma^2 r_f^2}{a^2 + \sigma^2} \right)\\
&= -\frac{1}{\lambda} (a^2 + \sigma^2) \frac{\sigma^2 r_f^2}{a^2 + \sigma^2}\left( \E[u^2|\pi^*_{T-2}]  +2\frac{a(r_f x-r_f^{-1}w)}{a^2 + \sigma^2}\E[u|\pi^*_{T-2}]  + (\frac{a(r_f x-r_f^{-1}w)}{a^2 + \sigma^2})^2\right)\\
&\quad+\frac{1}{2}\ln \left( \frac{a^2 + \sigma^2}{\pi \lambda} \frac{\sigma^2 r_f^2}{a^2 + \sigma^2}\right)\\
&=-\frac{1}{\lambda} \frac{\sigma^2 r_f^2}{a^2 + \sigma^2} \left( (a^2 + \sigma^2) \E[u^2|\pi^*_{T-2}]  +2a(r_f x-r_f^{-1}w) \E[u|\pi^*_{T-2}]  + \frac{a^2(r_f x-r_f^{-1}w)^2}{a^2 + \sigma^2}\right)\\
&\quad+\frac{1}{2}\ln \left( \frac{a^2 + \sigma^2}{\pi \lambda} \frac{\sigma^2 r_f^2}{a^2 + \sigma^2} \right).
\end{align*}
Thus, we can calculate %$J^*(T-2,x;w)$
\begin{align*}
&\quad J^*(T-2,x;w) \\
&=\frac{\sigma^2 r_f^2}{a^2 + \sigma^2} \left((a^2 + \sigma^2)  \E[u^2|\pi^*_{T-2}]\!+\!(2r_f x-2r_f^{-1}w) a \E[u|\pi^*_{T-2}] \right)\!+\!\frac{\sigma^2 r_f^2}{a^2 + \sigma^2}\left( r_f x\!-\!r_f^{-1}w \right)^2\\
& \quad +\lambda \int_{\R} \ln\pi^*_{T-2}(u)\pi^*_{T-2}(u) du+\frac{\lambda}{2} \ln \left( \frac{a^2 + \sigma^2}{\pi \lambda}\right)-(w-b)^2\\
&= - \frac{\sigma^2 r_f^2}{a^2 + \sigma^2} \frac{a^2(r_f x-r_f^{-1}w)^2}{a^2 + \sigma^2} + \frac{\sigma^2 r_f^2}{a^2 + \sigma^2} ( r_f x-r_f^{-1}w)^2 +\frac{\lambda}{2} \ln \left( \frac{a^2 + \sigma^2}{\pi \lambda} \frac{\sigma^2 r_f^2}{a^2 + \sigma^2} \right)  \\
&\quad +\frac{\lambda}{2} \ln \left( \frac{a^2 + \sigma^2}{\pi \lambda} \right)-(w-b)^2\\
&=(\frac{\sigma^2 r_f^2}{a^2 + \sigma^2})^2 (x-\rho_{T-2} w)^2 +\frac{\lambda}{2} \ln \left(\frac{\sigma^2 r_f^2}{a^2 + \sigma^2} \right)+\lambda \ln \left( \frac{a^2 + \sigma^2}{\pi \lambda} \right)-(w-b)^2.
\end{align*}

Now, we need to prove the optimal value function at period $t$ follows that
\begin{align*}
&\quad J^*(t,x;w) \\
&=\left(\frac{\sigma^2 r_f^2}{a^2 + \sigma^2} \right)^{T-t} ( x-\rho_t w)^2 +\frac{\lambda}{2} (T-t)\ln \left( \frac{a^2 + \sigma^2}{\pi \lambda} \right) +\frac{\lambda}{2}\sum_{i=t+1}^{T}(T-i)\ln \left( \frac{\sigma^2 r_f^2}{a^2 + \sigma^2} \right)\\
&\quad -(w-b)^2.
\end{align*}
where $\rho_t=( r_f^{-1})^{T-t}$, and the distribution of the optimal feedback control subjects to
\begin{align*}
\pi^{*}(u;t,x,w)
&= \mathcal{N} \left(u \Big|-\frac{ar_f( x-\rho_tw)}{a^2 + \sigma^2}, \frac{\lambda}{2(a^2 + \sigma^2)} \left(\frac{a^2 + \sigma^2}{\sigma^2 r_f^2} \right)^{T-t-1} \right).
\end{align*}

In fact, we proved the conjecture holds at period $T-1$ and $T-2$. Then we need to prove this conjecture at period $t$.

We first assume that this conjecture holds at period $t+1$. Thus, the value function at period $t$ can be transformed into the following equation
\begin{align*}
J(t,x;w) &=  \E \Big[\left(\frac{\sigma^2 r_f^2}{a^2 + \sigma^2} \right)^{T-t-1} ( x_{t+1}^{\pi}-\rho_{t+1} w)^2 +\frac{\lambda}{2}(T-t-1)\ln \left( \frac{a^2 + \sigma^2}{\pi \lambda} \right) \\
&\quad +\frac{\lambda}{2}\sum_{i=t+2}^{T}(T-i)\ln \left(  \frac{\sigma^2 r_f^2}{a^2 + \sigma^2} \right)+\lambda \int_{\R} \pi_{t}(u) \ln\pi_{t}(u) du \Big| x_{t}^{\pi}=x\Big]-(w-b)^2.
\end{align*}
Substituting the wealth updating process (\ref{2}) into this equation leads to 
\begin{align*}
&\quad J(t,x;w) \\
& =  \E \left[\left(\frac{\sigma^2 r_f^2}{a^2 + \sigma^2} \right)^{T-t-1} ( r_f x_{t}^{\pi} + r_{t}u_{t}^{\pi} - \rho_{t+1}w )^2+\lambda \int_{\R} \pi_{t}(u) \ln \pi_{t}(u) du \Big| x_{t}^{\pi}=x\right] \\
&\quad +\frac{\lambda}{2}(T-t-1)\ln \left( \frac{a^2 + \sigma^2}{\pi\lambda} \right) +\frac{\lambda}{2}\sum_{i=t+2}^{T}(T-i)\ln \left(\frac{\sigma^2 r_f^2}{a^2 + \sigma^2} \right)-(w-b)^2\\
& = \left(\frac{\sigma^2 r_f^2}{a^2 + \sigma^2} \right)^{T-t-1}  \E \Big[( r_f x_{t}^{\pi})^2  + (r_{t}u_{t}^{\pi})^2 + (\rho_{t+1}w)^2 + 2 r_f x_{t}^{\pi}r_{t}u_{t}^{\pi}-2\rho_{t+1}w r_f x_{t}^{\pi}\\ 
&\quad  -2\rho_{t+1}w r_{t}u_{t}^{\pi} \Big| x_{t}^{\pi}=x \Big]+ \lambda \int_{\R} \pi_{t}(u) \ln\pi_{t}(u) du +\frac{\lambda}{2}(T-t-1)\ln \left( \frac{a^2 + \sigma^2}{\pi \lambda} \right) \\
&\quad +\frac{\lambda}{2}\sum_{i=t+2}^{T}(T-i)\ln \left(  \frac{\sigma^2 r_f^2}{a^2 + \sigma^2} \right)-(w-b)^2\\
&= \left(\frac{\sigma^2 r_f^2}{a^2 + \sigma^2} \right)^{T-t-1} \Big[( r_f x)^2 + (a^2 + \sigma^2) \int_{\R} u^2 \pi_{t}(u) du + (\rho_{t+1}w)^2  -2\rho_{t+1}w r_f x\\
&\quad + 2r_f x a\int_{\R} u \pi_{t}(u) du- 2\rho_{t+1}w a\int_{\R} u \pi_{t}(u) du \Big] +\frac{\lambda}{2}(T-t-1) \ln \left( \frac{a^2 + \sigma^2}{\pi \lambda} \right) \\ 
&\quad+ \lambda \int_{\R} \pi_{t}(u) \ln\pi_{t}(u) du+\frac{\lambda}{2}\sum_{i=t+2}^{T}(T-i)\ln \left(  \frac{\sigma^2 r_f^2}{a^2 + \sigma^2} \right)-(w-b)^2\\
&= \int_{\R}\left( \left(\frac{\sigma^2 r_f^2}{a^2 + \sigma^2} \right)^{T-t-1} \left((a^2 + \sigma^2)  u^2 + (2r_f x-2\rho_{t+1}w) a u \right)+ \lambda  \ln\pi_{t}(u)\right) \pi_{t}(u) du \\
& \quad+\left(\frac{\sigma^2 r_f^2}{a^2 \!+\! \sigma^2} \right)^{T-t-1} \left(( r_f x)^2 \!+\! (\rho_{t+1}w)^2 -2\rho_{t+1}w r_f x\right)+\frac{\lambda}{2}(T-t-1)\ln \left( \frac{a^2 + \sigma^2}{\pi \lambda} \right) \\ 
&\quad+\frac{\lambda}{2}\sum_{i=t+2}^{T}(T-i)\ln \left(  \frac{\sigma^2 r_f^2}{a^2 + \sigma^2} \right)-(w-b)^2\\
&= \int_{\R}  \left( \left(\frac{\sigma^2 r_f^2}{a^2 + \sigma^2} \right)^{T-t-1}  \left((a^2 + \sigma^2)  u^2 + (2r_f x-2\rho_{t+1}w) a u \right)+ \lambda  \ln\pi_{t}(u)\right) \pi_{t}(u) du \\
& \quad+\left(\frac{\sigma^2 r_f^2}{a^2 + \sigma^2} \right)^{T-t-1} \left(r_f x- \rho_{t+1}w  \right)^2+\frac{\lambda}{2}(T-t-1)\ln \left( \frac{a^2 + \sigma^2}{\pi\lambda} \right)\\
&\quad +\frac{\lambda}{2}\sum_{i=t+2}^{T}(T-i)\ln \left(  \frac{\sigma^2 r_f^2}{a^2 + \sigma^2} \right)-(w-b)^2.
\end{align*}

We also minimize the value function at period $t$ about $\pi_{t}(u)$. Using the condition $\frac{\partial J(t,x;w)}{\partial \pi_{t}(u)} = 0$, we obtain that
\begin{align*}
\left(\frac{\sigma^2 r_f^2}{a^2 + \sigma^2} \right)^{T-t-1} \left((a^2 + \sigma^2)  u^2 + (2r_f x-2\rho_{t+1}w) a u \right)+ \lambda  \ln\pi_{t}(u) + \lambda =0.
\end{align*}
Applying the usual verification technique and using the fact that
$\pi \in \mathcal{P}(\R)$ if and only if
\begin{align*}
\int_{\R} \pi(u) du =1 \quad \mbox{ and } \quad \pi(u) \geq 0 \mbox{ a.e on } \R,
\end{align*}
we obtain a feedback control whose density function is given by
\begin{align*}
\pi^*(u;t,x,w) &= \frac{\exp \{-\frac{1}{\lambda} \left(\frac{\sigma^2 r_f^2}{a^2 + \sigma^2} \right)^{T-t-1}\left( (a^2 + \sigma^2)  u^2 + (2r_f x-2\rho_{t+1}w) a u \right) \}}{\int_{\R}\exp \{-\frac{1}{\lambda} \left(\frac{\sigma^2 r_f^2}{a^2 + \sigma^2} \right)^{T-t-1}\left( (a^2 + \sigma^2)  u^2 + (2r_f x-2\rho_{t+1}w) a u \right) \}du}\\
&= \mathcal{N} \left(u \Big|-\frac{ar_f (x-\rho_{t}w)}{a^2 + \sigma^2}, \frac{\lambda}{2(a^2 + \sigma^2)} \left(\frac{a^2 + \sigma^2}{\sigma^2 r_f^2} \right)^{T-t-1}\right).
\end{align*}
We also notice that
\begin{align*}
& \E[u|\pi^*_{T-2}] =-\frac{ar_f (x-\rho_{t}w)}{a^2 + \sigma^2},\\
& \E[u^2|\pi^*_{T-2}] = \left(\frac{ar_f (x-\rho_{t}w)}{a^2 + \sigma^2}\right) ^2 +  \frac{\lambda}{2(a^2 + \sigma^2)} \left(\frac{a^2 + \sigma^2}{\sigma^2 r_f^2} \right)^{T-t-1}.
\end{align*}
Substituting $\pi^*(u;t,x,w)$ into $J(t,x;w)$, we get 
\begin{align*}
&\quad J^*(t,x;w) \\
&= \int_{\R}  \left( \left(\frac{\sigma^2 r_f^2}{a^2 + \sigma^2} \right)^{T-t-1}  \left((a^2 + \sigma^2)  u^2 + (2r_f x-2\rho_{t+1}w) a u \right)+ \lambda  \ln\pi_{t}^*(u)\right) \pi_{t}^*(u) du \\
& \quad +\left(\frac{\sigma^2 r_f^2}{a^2 + \sigma^2} \right)^{T-t-1} \left(r_f x- \rho_{t+1}w  \right)^2+\frac{\lambda}{2}(T-t-1)\ln \left( \frac{a^2 + \sigma^2}{\pi \lambda} \right)\\
&\quad +\frac{\lambda}{2}\sum_{i=t+2}^{T}(T-i)\ln \left(  \frac{\sigma^2 r_f^2}{a^2 + \sigma^2} \right)-(w-b)^2\\
&=\left(\frac{\sigma^2 r_f^2}{a^2 + \sigma^2} \right)^{T-t-1} \left(\!(a^2 + \sigma^2)  \E[u^2|\pi^*_{t}]\! +\! 2(r_f x-\rho_{t+1}w) a \E[u|\pi^*_{t}]\! \right) + \lambda \int_{\R}\ln\pi^*_{t}(u)\pi^*_{t}(u)du\\
& \quad +\left(\frac{\sigma^2 r_f^2}{a^2 + \sigma^2} \right)^{T-t-1}\left( r_f x-\rho_{t+1}w \right)^2 +\frac{\lambda}{2}(T-t-1)\ln \left( \frac{a^2 + \sigma^2}{\pi \lambda} \right)\\
&\quad +\frac{\lambda}{2}\sum_{i=t+2}^{T}(T-i)\ln \left(  \frac{\sigma^2 r_f^2}{a^2 + \sigma^2} \right)-(w-b)^2.
\end{align*}
Then we obtain 
\begin{align*}
& \quad \int_{\R}  \ln\pi^*_{t}(u)\pi^*_{t}(u)du \\
&= -\frac{1}{\lambda} (a^2 + \sigma^2) \left(\frac{\sigma^2 r_f^2}{a^2 + \sigma^2} \right)^{T-t-1}\int_{\R} \left(\!u^2 \!+ \!2u\frac{a(r_f x- \rho_{t+1}w)}{a^2 + \sigma^2}\!+\! (\frac{a(r_f x-\rho_{t+1}w)}{a^2 + \sigma^2})^2 \!\right)\pi_{t}^*(u) du\\
&\quad +\frac{1}{2}\ln \left( \frac{a^2 + \sigma^2}{\pi \lambda}\left(\frac{\sigma^2 r_f^2}{a^2 + \sigma^2} \right)^{T-t-1} \right)\\
&= -\frac{1}{\lambda} (a^2 + \sigma^2) \left(\frac{\sigma^2 r_f^2}{a^2 + \sigma^2} \right)^{T-t-1}\left( \!\E[u^2|\pi^*_{t}]\! +\!2\frac{a(r_f x-\rho_{t+1}w)}{a^2 + \sigma^2}\E[u|\pi^*_{t}]\! +\! (\frac{a(r_f x-\rho_{t+1}w)}{a^2 + \sigma^2})^2\!\right)\\
&\quad+\frac{1}{2}\ln \left( \frac{a^2 + \sigma^2}{\pi \lambda} \left(\frac{\sigma^2 r_f^2}{a^2 + \sigma^2} \right)^{T-t-1}\right)\\
&=-\frac{1}{\lambda}\left(\frac{\sigma^2 r_f^2}{a^2 + \sigma^2} \right)^{T-t-1}\left( \!(a^2 + \sigma^2) \E[u^2|\pi^*_{t}]\!+\!2a(r_f x-\rho_{t+1}w) \E[u|\pi^*_{t}]\! +\! \frac{a^2(r_f x-\rho_{t+1}w)^2}{a^2 + \sigma^2}\!
\right)\\
&\quad+\frac{1}{2}\ln \left( \frac{a^2 + \sigma^2}{\pi \lambda}\left(\frac{\sigma^2 r_f^2}{a^2 + \sigma^2} \right)^{T-t-1}\right).
\end{align*}
Finally, we can calculate %$J^*(t,x;w)$
\begin{align*}
&\quad J^*(t,x;w) \\
&=\left(\frac{\sigma^2 r_f^2}{a^2 + \sigma^2} \right)^{T-t-1} \Big((a^2 + \sigma^2)  \E[u^2|\pi^*_{t}] + (2r_f x-2\rho_{t+1}w) a \E[u|\pi^*_{t}] \Big) \\
& \quad + \lambda\int_{\R}\ln\pi^*_{t}(u)\pi^*_{t}(u) du+ \left(\frac{\sigma^2 r_f^2}{a^2 + \sigma^2} \right)^{T-t-1}\left( r_f x-\rho_{t+1}w \right)^2 \\
&\quad +\frac{\lambda}{2}(T-t-1)\ln \left( \frac{a^2 + \sigma^2}{\pi \lambda} \right)+\frac{\lambda}{2}\sum_{i=t+2}^{T}(T-i) \ln \left(  \frac{\sigma^2 r_f^2}{a^2 + \sigma^2} \right)-(w-b)^2\\
&=- \left(\frac{\sigma^2 r_f^2}{a^2 + \sigma^2} \right)^{T-t-1}  \frac{a^2(r_f x-\rho_{t+1}w)^2}{a^2 + \sigma^2}  + \left(\frac{\sigma^2 r_f^2}{a^2 + \sigma^2} \right)^{T-t-1}\left( r_f x-\rho_{t+1}w \right)^2\\
&\quad +\frac{\lambda}{2}(T-t-1) \ln \left( \frac{a^2 + \sigma^2}{\pi \lambda} \right)+\frac{\lambda}{2}\sum_{i=t+2}^{T}(T-i)\ln \left(  \frac{\sigma^2 r_f^2}{a^2 + \sigma^2} \right) \\
&\quad +\frac{\lambda}{2}\ln \left( \frac{a^2 + \sigma^2}{\pi \lambda}\left(\frac{\sigma^2 r_f^2}{a^2 + \sigma^2} \right)^{T-t-1}\right)-(w-b)^2\\
&=\left(\frac{\sigma^2 r_f^2}{a^2 + \sigma^2} \right)^{T-t-1} \frac{\sigma^2}{a^2 + \sigma^2} (r_f x-\rho_{t+1}w)^2 +\frac{\lambda}{2}(T-t-1) \ln \left( \frac{a^2 + \sigma^2}{\pi \lambda} \right)\\
&\quad+\frac{\lambda}{2}\sum_{i=t+2}^{T}(T-i)\ln \left(  \frac{\sigma^2 r_f^2}{a^2 + \sigma^2} \right)+\frac{\lambda}{2}\ln \left( \frac{a^2 + \sigma^2}{\pi \lambda}\left(\frac{\sigma^2 r_f^2}{a^2 + \sigma^2} \right)^{T-t-1}\right)-(w-b)^2\\
&=\left(\frac{\sigma^2 r_f^2}{a^2 + \sigma^2} \right)^{T-t} ( x-\rho_{t}w)^2 +\frac{\lambda}{2}(T-t)\ln \left( \frac{a^2 + \sigma^2}{\pi \lambda} \right)+\frac{\lambda}{2}\sum_{i=t+1}^{T}(T-i)\ln \left(  \frac{\sigma^2 r_f^2}{a^2 + \sigma^2} \right)\\
&\quad -(w-b)^2.
\end{align*}
We complete the proof of Theorem 1.

\section{Proof of Theorem 2}
From the initial value function, we calculate
\begin{align*}
&\quad J^{\pi^0}(t,x;w)\\
&=\E \left[J^{\pi^0}(t+1,x_{t+1};w)+\lambda \int_{\R}\pi_t^0(u) \ln \pi_t^0(u) du \Big| x_t = x \right]\\
&=\E \left[A^{T-t-1}(x_{t+1}-\rho_{t+1}w)^2+\lambda \int_{\R}\pi_t^0(u) \ln \pi_t^0(u) du \Big| x_t = x \right]+f(t+1)\\
&=\E \left[A^{T-t-1}(r_f x_t+r_t u_t -\rho_{t+1}w)^2+\lambda \int_{\R}\pi_t^0(u) \ln \pi_t^0(u) du \Big| x_t = x \right]+f(t+1)\\
&=\int_{\R}\left[A^{T-t-1}((2r_f x-2\rho_{t+1}w)au+(a^2 + \sigma^2)u^2)+\lambda \ln \pi_t^0(u) \right]\pi_t^0(u)du\\
& \quad +A^{T-t-1}(r_f x-\rho_{t+1}w)^2+f(t+1).
\end{align*}

We can use the condition $\pi_t^{1}(u;x,w) = \arg \mathop{\min}\limits_{\pi^0_t(u)} J^{\pi^0}(t,x;w)$ to make one iteration. We let $\frac{\partial J^{\pi^0}(t,x;w)}{ \partial \pi_{t}^0(u) } = 0 $, then we can get
\begin{align*}
A^{T-t-1}\left[(2r_f x-2\rho_{t+1}w)au+(a^2 + \sigma^2)u^2\right]+\lambda \ln \pi_t^0(u)+\lambda = 0.
\end{align*}

Applying the usual verification technique and using the fact that
$\pi \in \mathcal{P}(\R)$ if and only if
\begin{align*}
\int_{\R} \pi_t^0(u) du =1 \quad \mbox{ and } \quad \pi_t^0(u) \geq 0 \mbox{ a.e on } \R,
\end{align*}
we obtain the feedback control $\pi_t^1(u;x,w)$ whose density function is given by
\begin{align*}
\pi_t^1(u;x,w) &= \frac{\exp \{-\frac{1}{\lambda} A^{T-t-1} \left[ (a^2 + \sigma^2)  u^2 + (2r_f x-2\rho_{t+1}w) a u \right] \}}{\exp \{-\frac{1}{\lambda} A^{T-t-1} \left[ (a^2 + \sigma^2)  u^2 + (2r_f x-2\rho_{t+1}w) a u \right] \}}\\
&= \mathcal{N} \left(u \Big|-\frac{a r_f (x-\rho_t w)}{a^2 + \sigma^2}, \frac{\lambda}{2(a^2 + \sigma^2)A^{T-t-1}} \right).
\end{align*}
Then we calculate
\begin{align*}
&\quad \int_{\R}\ln \pi_t^1(u) \pi_t^1(u)du\\
&=\int_{\R}\ln \left(\left[\frac{2\pi \lambda}{2(a^2+\sigma^2)A^{T-t-1}} \right]^{-\frac{1}{2}}\exp \left\{-\frac{(a^2+\sigma^2)A^{T-t-1}}{\lambda}\left[u+\frac{a r_f(x-\rho_t w)}{a^2+\sigma^2} \right]^2 \right\}\right) \pi_t^1(u)du\\
&=-\frac{(a^2+\sigma^2)A^{T-t-1}}{\lambda}\left[\E(u^2|\pi_t^1(u))+\frac{2a r_f (x-\rho_t w)}{a^2+\sigma^2}\E(u|\pi_t^1(u))+\frac{a^2 r_f^2 (x-\rho_t w)^2}{(a^2+\sigma^2)^2} \right]\\
& \quad +\frac{1}{2}\ln \left( \frac{(a^2+\sigma^2)A^{T-t-1}}{\pi \lambda}\right).
\end{align*}
Thus, we can get the corresponding value function $J^{\pi^1}(t,x;w)$
\begin{align*}
&\quad J^{\pi^1}(t,x;w)\\
&=A^{T-t-1}(r_f x-\rho_{t+1}w)^2-A^{T-t-1}\frac{a^2(r_f x-\rho_{t+1}w)^2}{a^2+\sigma^2}+\frac{\lambda}{2}\ln \frac{a^2+\sigma^2}{\pi \lambda}\\
&\quad +\frac{\lambda \ln A}{2}(T-t-1)+f(t+1)\\
&=A^{T-t-1}\frac{\sigma^2 r_f^2}{a^2+\sigma^2}(x-\rho_t w)^2+\frac{\lambda}{2}\ln \frac{a^2+\sigma^2}{\pi \lambda}+\frac{\lambda \ln A}{2}(T-t-1)+f(t+1).
\end{align*}
Also, we can get the following result 
\begin{align*}
J^{\pi^1}(t,x;w) \leq J^{\pi^0}(t,x;w).
\end{align*}

After updating our policy for $j$ times, we can calculate $J^{\pi^j}(t,x;w)$
\begin{align*}
&\quad J^{\pi^j}(t,x;w)\\
&=A^{T-t-j}(\frac{\sigma^2 r_f^2}{a^2+\sigma^2})^j (x-\rho_t w)^2+\frac{\lambda}{2} j \ln(\frac{a^2+\sigma^2}{\pi \lambda})+\frac{\lambda}{2} \sum_{i=0}^{j-1} i\ln(\frac{\sigma^2 r_f^2}{a^2+\sigma^2})\\
& \quad +\frac{\lambda \ln A}{2} j(T-t-j)+f(t+j).
\end{align*}
Next, we derive the $(j+1)$-th iteration
\begin{align*}
&\quad J^{\pi^{j}}(t,x;w)\\
&=\E \left[J^{\pi^{j}}(t+1,x_{t+1};w)+\lambda \int_{\R}\pi_t^j(u) \ln \pi_t^j(u)du \Big| x_t = x \right]\\
&=\E \left[A^{T-t-1-j}(\frac{\sigma^2 r_f^2}{a^2+\sigma^2})^j (x_{t+1}-\rho_{t+1} w)^2+\lambda \int_{\R}\pi_t^j(u) \ln \pi_t^j(u)du \Big| x_t = x \right]\\
&\quad +\frac{\lambda}{2} j \ln(\frac{a^2+\sigma^2}{\pi \lambda})+\frac{\lambda}{2} \sum_{i=0}^{j-1} i\ln(\frac{\sigma^2 r_f^2}{a^2+\sigma^2})+\frac{\lambda \ln A}{2} j(T-t-1-j)+f(t+1+j)\\
&=\E \left[A^{T-t-1-j}(\frac{\sigma^2 r_f^2}{a^2+\sigma^2})^j (r_f x_t+u_t r_t-\rho_{t+1} w)^2+\lambda \int_{\R}\pi_t^j(u) \ln \pi_t^j(u)du \Big| x_t = x \right]\\
&\quad +\frac{\lambda}{2} j \ln(\frac{a^2+\sigma^2}{\pi \lambda})+\frac{\lambda}{2} \sum_{i=0}^{j-1} i\ln(\frac{\sigma^2 r_f^2}{a^2+\sigma^2})+\frac{\lambda \ln A}{2} j(T-t-1-j)+f(t+1+j)\\
&=\int_{\R} \left[A^{T-t-1-j}(\frac{\sigma^2 r_f^2}{a^2+\sigma^2})^j ((2r_f x-2\rho_{t+1}w)au+(a^2+\sigma^2)u^2) +\lambda \ln \pi_t^j(u) \right]\pi_t^j(u)du\\
&\quad +\frac{\lambda}{2} j \ln(\frac{a^2+\sigma^2}{\pi \lambda})+A^{T-t-1-j}(\frac{\sigma^2 r_f^2}{a^2+\sigma^2})^j(r_f x-\rho_{t+1}w)^2+\frac{\lambda}{2} \sum_{i=0}^{j-1} i\ln(\frac{\sigma^2 r_f^2}{a^2+\sigma^2})\\
&\quad +\frac{\lambda \ln A}{2} j(T-t-1-j)+f(t+1+j).
\end{align*}
Then we utilize the condition $\pi_t^{j+1}(u;x,w) = \arg \mathop{\min}\limits_{\pi^j_t(u)} J^{\pi^j}(t,x;w)$ to derive one iteration. Let $\frac{\partial J^{\pi^j}(t,x;w)}{ \partial \pi_{t}^j(u) } = 0$, then we can get
\begin{align*}
A^{T-t-1-j}(\frac{\sigma^2 r_f^2}{a^2+\sigma^2})^j\left[(2r_f x-2\rho_{t+1}w)au+(a^2 + \sigma^2)u^2\right]+\lambda \ln \pi_t^j(u)+\lambda = 0.
\end{align*}
Applying the usual verification technique and using the fact that
$\pi \in \mathcal{P}(\R)$ if and only if
\begin{align*}
\int_{\R} \pi_t^j(u) du =1 \quad \mbox{ and } \quad \pi_t^j(u) \geq 0 \mbox{ a.e on } \R,
\end{align*}
we obtain a feedback control $\pi_t^{j+1}(u;x,w)$ whose density function is given by
\begin{align*}
\pi_t^{j+1}(u;x,w)&= \mathcal{N} \left(u \Big|-\frac{a r_f (x-\rho_t w)}{a^2 + \sigma^2}, \frac{\lambda}{2(a^2 + \sigma^2)A^{T-t-1-j}}(\frac{a^2+\sigma^2}{\sigma^2 r_f^2})^j \right).
\end{align*}
Then we calculate
\begin{align*}
&\quad \int_{\R}\ln \pi_t^{j+1}(u) \pi_t^{j+1}(u)du\\
&=-\int_{\R}\frac{(a^2+\sigma^2)A^{T-t-1-j}}{\lambda}(\frac{\sigma^2 r_f^2}{a^2+\sigma^2})^j \left[u^2\!+\!\frac{2a r_f(x-\rho_t w)}{a^2+\sigma^2}u\!+\!\frac{a^2 r_f^2 (x-\rho_t w)^2}{(a^2+\sigma^2)^2}\!\right] \pi_t^{j+1}(u)du\\
&\quad +\frac{1}{2}\ln(\frac{a^2+\sigma^2}{\pi \lambda})+\frac{\ln A}{2}(T-t-1-j)+\frac{j}{2}\ln(\frac{\sigma^2 r_f^2}{a^2+\sigma^2}).
\end{align*}
Thus, we get $J^{\pi^{j+1}}(t,x;w)$
\begin{align*}
&\quad J^{\pi^{j+1}}(t,x;w)\\
&=-A^{T-t-1-j}(\frac{\sigma^2 r_f^2}{a^2+\sigma^2})^j \frac{a^2 r_f^2 (x-\rho_t w)^2}{a^2+\sigma^2}+\frac{\lambda}{2}\ln(\frac{a^2+\sigma^2}{\pi \lambda})+\frac{\lambda \ln A}{2}(T-t-1-j)\\
&\quad +\frac{\lambda}{2} j \ln(\frac{\sigma^2 r_f^2}{a^2+\sigma^2})+\frac{\lambda}{2} j \ln(\frac{a^2+\sigma^2}{\pi \lambda})+A^{T-t-1-j}(\frac{\sigma^2 r_f^2}{a^2+\sigma^2})^j(r_f x-\rho_{t+1}w)^2\\
&\quad +\frac{\lambda}{2} \sum_{i=0}^{j-1} i\ln(\frac{\sigma^2 r_f^2}{a^2+\sigma^2})+\frac{\lambda \ln A}{2} j(T-t-1-j)+f(t+1+j)\\
&=A^{T-t-1-j}(\frac{\sigma^2 r_f^2}{a^2+\sigma^2})^{j+1}(x-\rho_t w)^2+\frac{\lambda}{2} (j+1) \ln(\frac{a^2+\sigma^2}{\pi \lambda})+\frac{\lambda}{2} \sum_{i=0}^{j} i\ln(\frac{\sigma^2 r_f^2}{a^2+\sigma^2})\\
&\quad +\frac{\lambda \ln A}{2} (j+1)(T-t-1-j)+f(t+1+j).
\end{align*}
Also, we can get the following result
\begin{align*}
J^{\pi^{j+1}}(t,x;w) \leq J^{\pi^j}(t,x;w).
\end{align*}

In fact, after deriving $T-t$ iterations, $\pi^{T-t}(u;x,w)$ will converge to the $\pi^*(u;x,w)$
\begin{align*}
\pi_t^{T-t}(u;x,w)&= \mathcal{N} \left(u \Big|-\frac{a r_f (x-\rho_t w)}{a^2 + \sigma^2}, \frac{\lambda}{2(a^2 + \sigma^2)}(\frac{a^2+\sigma^2}{\sigma^2 r_f^2})^{T-t-1} \right)\\
&=\pi^*_t(u;x,w).
\end{align*}
We utilize the fact that $f(t+T-t)=f(T)=-(w-b)^2$ to prove that $J^{\pi^{T-t}}(t,x;w)$ will converge to the $J^*(t,x;w)$
\begin{align*}
&\quad J^{\pi^{T-t}}(t,x;w)\\
&=(\frac{\sigma^2 r_f^2}{a^2+\sigma^2})^{T-t}(x-\rho_t w)^2\!+\!\frac{\lambda}{2} (T-t) \ln(\frac{a^2+\sigma^2}{\pi \lambda})\!+\!\frac{\lambda}{2} \sum_{i=0}^{T-t-1} i\ln(\frac{\sigma^2 r_f^2}{a^2+\sigma^2})+f(T)\\
&=(\frac{\sigma^2 r_f^2}{a^2+\sigma^2})^{T-t}(x-\rho_t w)^2\!+\!\frac{\lambda}{2} (T-t) \ln(\frac{a^2+\sigma^2}{\pi \lambda})\!+\!\frac{\lambda}{2}\sum_{i=t+1}^{T}(T-i)\ln ( \frac{\sigma^2 r_f^2}{a^2 + \sigma^2} )+f(T)\\
&=(\frac{\sigma^2 r_f^2}{a^2+\sigma^2})^{T-t}(x-\rho_t w)^2\!+\!\frac{\lambda}{2} (T-t) \ln(\frac{a^2+\sigma^2}{\pi \lambda})\!+\!\frac{\lambda}{2}\sum_{i=t+1}^{T}(T-i)\ln ( \frac{\sigma^2 r_f^2}{a^2 + \sigma^2} )-(w-b)^2\\
&=J^*(t,x;w).
\end{align*}
We complete the proof of Theorem 2. 

\newpage
\bibliographystyle{apalike}
\bibliography{ref}

\end{document}